\documentclass[fleqn,usenatbib]{mnras}

\usepackage{newtxtext,newtxmath}

\usepackage[T1]{fontenc}
\DeclareRobustCommand{\VAN}[3]{#2}
\let\VANthebibliography\thebibliography
\def\thebibliography{\DeclareRobustCommand{\VAN}[3]{##3}\VANthebibliography}

\usepackage{graphicx}   
\usepackage{amsmath}    
\usepackage{multirow}
\usepackage[table,xcdraw]{xcolor}
\usepackage{arydshln}

\title[G and NG Clusters PPS Analysis]{Investigating the Projected Phase Space of Gaussian and Non-Gaussian Clusters}

\author[Sampaio et al. 2020]{V. M. Sampaio,$^{1,2}$\thanks{E-mail: vitorms999@gmail.com}
R. R. de Carvalho ,$^{2}$ I. Ferreras,$^{3,4,5}$ T. F. Lagan\'a,$^{2}$
\newauthor A. L. B. Ribeiro,$^{6}$ S. B. Rembold$^{7}$
\\
$^{1}$ Divis\~ao de Astrof\'\i sica (INPE-MCT), S\~ao Jos\'e dos Campos, 12227-010, SP, Brazil\\
$^{2}$ NAT - Universidade Cruzeiro do Sul /  Universidade Cidade de S\~ao Paulo, 01506-000, SP, Brazil\\
$^{3}$ Instituto de Astrof\'isica de Canarias, Calle V\'i a L\'actea s/n,
E38205, La Laguna, Tenerife, Spain\\
$^{4}$ Department of Physics and Astronomy, University College London, Gower Street, London WC1E 6BT, UK\\
$^{5}$ Departamento de Astrof\'\i sica, Universidad de La Laguna, E38206 La Laguna, Tenerife, Spain\\
$^{6}$ Laborat\'orio de Astrof\'\i sica Te\'orica e Observacional, Universidade Estadual de Santa Cruz, 45650-000, BA, Brazil\\
$^{7}$ Universidade Federal de Santa Maria, 97105-900, RS, Brazil
}

\date{Accepted 2021 March 03. Received 2021 March 02; in original form 2020 December 01}

\pubyear{2020}

\begin{document}
\label{firstpage}
\pagerange{\pageref{firstpage}--\pageref{lastpage}}
\maketitle

\begin{abstract}
By way of the projected phase-space (PPS), we investigate the relation
between galaxy properties and cluster environment in a subsample of
groups from the Yang Catalog. The sample is split according to the
gaussianity of the velocity distribution in the group into gaussian (G) and non-gaussian (NG). Our sample is limited to massive clusters
with $\rm M_{200} \geq 10^{14} M_{\odot}$ and $\rm 0.03\leq z \leq 0.1$.
NG clusters are more massive, less concentrated and have an
excess of faint galaxies compared to G clusters. NG clusters show
mixed distributions of galaxy properties in the PPS compared to the G
case. Using the relation between infall time and locus on the PPS, we
find that, on average, NG clusters accreted $\rm \sim 10^{11}\,M_{\odot}$
more stellar mass in the last $\sim 5$\,Gyr than G clusters. The 
relation between galaxy properties and infall time is significantly
different for galaxies in G and NG systems. The more mixed distribution
in the PPS of NG clusters translates into shallower relations with 
infall time. Faint galaxies whose first crossing of the cluster virial radius happened 2-4 Gyr ago in NG clusters are older and more metal-rich than in G systems. All these results suggest that NG clusters experience a higher accretion of pre-processed galaxies, which characterizes G and NG clusters as different environments to study galaxy evolution.
\end{abstract}




\section{Introduction}
\label{sec:Introduction}

The study of galaxies in the nearby Universe reveals that their
evolution is fundamentally dependent on their environment. From
the core to the outskirts of clusters, the galactic density roughly
spans seven orders of magnitude, and the morphology-density relation
established by \citet{Dressler} is one of the simplest examples of
how the environment affects galaxy properties. Early-type galaxies
(ETGs) inhabit the core (high-density regime), while Late-type
galaxies (LTGs) preferentially populate the outskirts (i.e. at 
lower densities). In the last decades, observations show that galaxies
usually evolve from
star-forming late-type systems to passive/quiescent early-type ones. These two extreme cases explain the bimodality observed, for
instance, in star-formation rate (SFR,
e.g. \citealt{wetzel2012galaxy}) and in gas content. The dichotomy
between star-forming LTGs and quiescent ETGs in the stellar mass
versus SFR defines the so-called Blue Cloud (BC) and Red Sequence
(RS), respectively. An intermediate region, Green Valley (GV),
reinforces the idea of having physical mechanisms that quench star
formation transforming a system in the BC into one in the RS
(e.g. \citealt{2019MNRAS.488L..99A}). In clusters,
galaxies lose their galactic hot gas as they fall (depending
on their total mass), even before entering the cluster virial region
(through strangulation/starvation, \citealt{1980ApJ...237..692L,2000ApJ...540..113B,2017MNRAS.466.3460V}). After
crossing the virial region, gravitational tides from the cluster deep
potential well strip away the interstellar medium, stars and dark
matter from the infalling galaxy (tidal stripping, \citealt{1999MNRAS.302..771J,2006MNRAS.366..429R}).
Furthermore, the hot gas in the
intra-cluster medium (ICM) exerts pressure on galaxies moving
within the cluster and may remove gas via ram
pressure stripping \citep[RPS,][]{1972ApJ...176....1G,1999MNRAS.308..947A}. Galaxy clusters also provide a suitable environment for galaxy-galaxy interactions, especially in the core. Indirect continuous encounters between galaxies within the
cluster may leave interacting galaxies with distorted features. Direct
encounters may lead to galaxy mergers, and cause a burst in star
formation over a short time scale and rapidly exhaust the gas
component \citep{2005ApJ...622L...9S,2008MNRAS.384..386C,2010ApJ...720L.149T}. In addition, previous episodes inhabiting groups with lower halo mass can alter the properties of galaxies even before infalling into clusters, an effect called
pre-processing \citep{2004PASJ...56...29F,2013MNRAS.431L.117M,2019A&A...632A..49S}. There are also internal feedback processes driven by AGN, supernovae and stellar winds that cause gas outflows that diminish the gas reservoir of galaxies (mass
quenching, \citealt{1974MNRAS.169..229L,1986ApJ...303...39D,Bongiorno}).

The variety of mechanisms quenching star formation fuels the debate
concerning the most important driver, and the related time scales.
\cite{2020MNRAS.491.5406T} study galaxies in the local
Universe and suggest that quenching has an extended phase ($\sim 5$\,Gyr)
of starvation. However, several 
studies find that the main quenching mechanism is dependent on halo
mass \citep{2016MNRAS.457.4360Z}. \cite{2010ApJ...721..193P} show that
a stellar mass related mechanism plays a major role in quenching massive
galaxies. \cite{2020MNRAS.491.5406T} show also that gas outflows are
increasingly relevant at decreasing stellar mass. In addition,
\cite{2017ApJ...843..128R} (R17, hereafter) suggest that galaxies in clusters lose a
significant fraction of their mass over long time scales due to tidal
mass loss (TML, see their Fig.~3). On the
other hand, in short time-scales ($\sim 2$ Gyr), RPS is found to be the 
main quenching mechanism, but only above a density threshold
(\citealt{2019ApJ...873...42R}, R19 hereafter). The cumulative effect
of different quenching mechanisms acting simultaneously in galaxies results
in a non-linearity of the general quenching process. Several models have been
proposed in the last decades to simplify this inherent non-linearity.
One of the most recently
adopted hypothesis is the so-called ``delayed-then-rapid'' quenching model \citep{2013MNRAS.432..336W}. In
this model, an infalling galaxy is unaffected by the cluster
environment for a delay time after which it becomes a satellite and is mostly
quenched due to starvation in this phase. After the delay time, the 
cluster environment strongly affects star formation, that is rapidly quenched due to RPS.

Star formation quenching is also conditional on cluster 
properties. For instance, RPS depends on the ICM density and
its velocity relative to the galaxy. In order to characterize different environments,
substructure analyses in the optical
(e.g. \citealt{1988AJ.....95..985D,1997ApJ...482...41G}) and X-ray
(e.g. \citealt{2001A&A...378..408S,2009ApJ...699.1178Z}) indicate that
many clusters are not fully virialized. The degree of relaxation is
also related to the orbital parameters. Infalling galaxies have highly radial
orbits in the outskirts, while virialized objects show circular orbits
within the virial radius. It follows from statistical mechanics that
the equilibrium state of a dynamical system may be well described by a
defined velocity distribution. \cite{1957AZh....34..770O} and
\cite{1967MNRAS.136..101L} suggest a Maxwell-Boltzmann distribution in
velocity as an equilibrium state for gravitationally bound
systems. This result, however, rests on simplifying assumptions that may be
unrealistic (clusters evolving in isolation and gravitation as
the only interaction, for example). An additional limitation is that 
the observations are projected along the line-of-sight
(LOS). The use of N-body simulations gives support to the adoption of
a Maxwell-Boltzmann distribution, that leads to a Gaussian distribution
in velocity projected along the line of sight
\citep{merrall2003relaxation,2005NewA...10..379H}. Previous studies
investigate the difference between clusters with projected velocity
distributions well-fit by a Gaussian (G) and those with a non-Gaussian
(NG) velocity profile and find that: 1) NG clusters have an excess of
star-forming galaxies \citep{2010MNRAS.409L.124R}; 2) the stellar
population parameters of infalling and virialized galaxies in NG clusters
are not well separated as in G clusters \citep{2013MNRAS.434..784R};
3) there is evidence of a higher infall rate of pre-processed galaxies
in NG clusters \citep{2017MNRAS.467.3268R,2017AJ....154...96D}; 4) the
velocity dispersion profiles of cluster members is significantly 
different between G and NG systems \citep{2018MNRAS.473L..31C}; and 5)
simulations show that NG systems suffered their last major merger more
recently than G systems \citep{2019MNRAS.490..773R}. These results
point towards a higher infall rate in NG clusters in comparison to G
clusters. However, the separation between G and NG clusters rely on a
robust measure for gaussianity. Anderson-Darling test
\citep{anderson1952asymptotic} and Gaussian Mixture Models
\citep{reynolds2009gaussian} are among the most employed methods to
measure gaussianity. However, comparing two general distributions is a longstanding problem in statistics. For instance, several methods implement the testing of the null hypothesis that two populations corresponding to different datasets originate from the same parent distribution (e.g. Mann–Whitney–Wilcoxon, Hodges–Lehmann, Kruskal–Wallis, etc.) but none yields a definite answer to the problem \citep{2012msma.book.....F}. In this particular case, the aim is for a clear, quantitative and unambiguous distinction between the velocity distributions into G and NG.

As an option to study the velocity distribution, the Lagrangian formulation
of motion enables the study of the dynamical evolution of physical
systems through the ``phase-space'', which combines both spatial and
velocity coordinates into a single space. In clusters, this complex
6D space is simplified to a diagram comprising cluster-centric distance and
velocity. Galaxies infalling in clusters have a well-defined trajectory in 
phase-space (see Fig.~1 of R17).
Nevertheless, observations are limited to projected quantities and so
phase space is constrained to a projected version, the so-called projected
phase space (PPS, hereafter). In the PPS, the aforementioned G or NG
velocity distributions are simply a projection along the y-axis. The PPS
is thus a more informative space than velocity distribution alone.
It is also important to emphasize that environmental effects,
such as RPS, are conditional on the velocity of the infalling galaxy and
thus the PPS provides a more suitable way to study environmental effects.
\cite{2020MNRAS.tmp.3623O} combine data from the SDSS to
show that galaxies take around $\sim 3$\,Gyr after the first
pericentric passage to be quenched. Although the well-defined
trajectory in phase space becomes degenerate in the PPS due to
projection effects, \cite{2011MNRAS.416.2882M} and
\cite{2013MNRAS.431.2307O} use numerical simulations and suggest that galaxies at different orbits -- and consequently in different dynamical stages inside the cluster -- 
occupy distinct regions in the PPS. \cite{2019MNRAS.484.1702P} (hereon P19) study the
distribution of infall time ($\rm t_{inf}$) -- defined as the time duration from when the galaxy reached the virial radius of the main progenitor of its present-day host environment, for the first time -- in the PPS and define regions
that constrain galaxies within a narrow width of $\rm t_{inf}$.
R17 investigate the relation between
tidal mass loss, $\rm t_{inf}$ and the region occupied in the
PPS. Finally, \cite{2020ApJS..247...45R} propose a relation between
SFR and time since infall based on the region occupied in the PPS.

In this work, we follow a methodology that focuses on the PPS to investigate
further differences between G and NG clusters. We map the PPS with a grid and
explore the properties in each galactic environment. Taking stellar mass
as a key parameter, we derive a rough estimate of
the infall rate in NG clusters from the observed distribution. We
perform a statistical study in the different regions and evaluate the
global properties of the PPS of G and NG clusters. We also build a
relation between galaxy properties and infall time using PPS regions
presented in P19.

This paper is organized as follows: in $\S$~2 we define the sample,
present the stellar population parameters and dynamical properties of
the galaxies, and describe the related methods. In $\S$~3 we
present a first characterization of the structure of G and NG clusters
and their galactic content. $\S$~4 introduces the methods adopted 
to build the PPS and define different loci of interest. In $\S$~5
we connect galaxy properties with their position on the PPS
diagram. In $\S$~6 we explore the relation between time since
infall and stellar population parameters. In $\S$~7 we derive a
first estimate of the infall rate in NG clusters with respect to G
clusters. Finally, in $\S$~8 we discuss the relationship between the gaussianity
of the velocity distribution and the properties of cluster members.
This paper adopts a flat $\rm \Lambda CDM$ cosmology with
$[\rm \Omega_{M},\Omega_{\Lambda},H_{0}] = [0.27,0.73,72$ km $\rm s^{-1} Mpc^{-1}]$.  \renewcommand{\arraystretch}{1.5}

\section{Sample and Data}
\label{sec:Sample}

Our sample is based on the Yang Catalog \citep{2007ApJ...671..153Y},
which uses a halo finder algorithm applied to the New York University
Value-Added Galaxy Catalog (NYU - VAGC,
\citealt{2005AJ....129.2562B}). The catalog was originally based on the fifth
data release of the Sloan Digital Sky Survey (SDSS-DR5,
\citealt{2007ApJS..172..634A}). We make use of an updated version, 
presented in \cite{2017AJ....154...96D} (dC17, hereafter), based on the seventh data release (SDSS-DR7).

\subsection{Data Selection and Membership Assignment}
\label{subsec:Data_Select}

We use similar data to those presented in dC17 and
\citet{2018MNRAS.473L..31C}. We select SDSS-DR7 galaxies with
line-of-sight velocities in the range $ \rm \pm$ 4,000 km\,s$^{-1}$
and a projected distance $\rm d_{proj}\leq 2.5h^{-1}$Mpc
(i.e. 3.47\,Mpc for h = 0.72) from the clustercentric coordinates described
in the Yang Catalog. We use galaxies in the redshift interval
0.03 $\leq$ z $\leq$ 0.1, and r-band magnitudes $\rm m_{r} \leq 17.78$, which is the survey
spectroscopic completeness limit. These criteria guarantee that we
probe the luminosity function to $\rm M^{\star} + 1$\,mag. The
lower limit in redshift avoids large aperture-related bias in the stellar population
parameters due to the fixed 3\,arcsec diameter of the SDSS
fibers.

Galaxy membership is defined via an iterative Shiftgapper technique
(see \citealt{2009MNRAS.399.2201L}). Next we briefly describe how this technique works. First, we select SDSS-DR7 galaxies around the cluster center presented in the Yang Catalog \footnote{We highlight this is the only use of the Yang Catalog.} to feed the Shiftgapper technique. Then the algorithm follows the methodology presented in \cite{1996ApJ...473..670F}. Namely, we apply a gap technique in radial bins with sizes $\rm 0.43h^{-1}$ Mpc\footnote{This bin size ensures at least 15 galaxies in each bin} and remove galaxies with a velocity gap greater than 1000 $km\,s^{-1}$ relative to the mean cluster velocity (see Fig.~1 in \citealt{1996ApJ...473..670F}). This procedure is reiterated until there are no more interlopers. We define the clustercentric coordinates as the luminosity weighted RA, DEC and median redshift. We use this technique to redefine the Yang catalog due to two main reasons: 1) Shiftgapper avoids prior
assumptions about the dynamical state of the system; and 2) comparison shows that Shiftgapper is less restrictive than the Halo mass finder algorithm, which is relevant to works investigating satellite galaxy properties. We perform a virial
analysis (see \citealt{2009MNRAS.392..135L}) in the final list of
members to estimate dynamical quantities like virial radius ($\rm
R_{200}$), virial mass ($\rm M_{200}$), and velocity dispersion along
the line of sight ($\rm \sigma_{LOS}$). We highlight that the
Shiftgapper method returns a list of galaxies without interlopers, which will
be relevant in Section \ref{sec:infall_rate_estimate}. Finally, we
restrain our sample to systems with at least 20 galaxy members within
$\rm R_{200}$ (see Section \ref{subsec:Gaussianity_Measure} for this
threshold explanation), resulting in 319 systems. We separate member
galaxies into two different luminosity regimes: 1) Bright (B): 
$0.03 \leq z \leq 0.1$ and $\rm M_{r} \leq -20.55 \sim M^{\star} + 1$\,mag,
where $\rm M_{r}$ is the limiting absolute magnitude in the r-band; and 2)
Faint (F): $0.03 \leq z \leq 0.04$ and
$\rm -20.55 < M_{r} \leq -18.40 \sim M^{\star} + 3$\,mag.
The redshift upper limit in the Faint sample 
corresponds to the spectroscopic completeness limit for $\rm M_{r} = -18.40$ in the SDSS.

\subsection{Characterizing Different Galactic Environments}
\label{subsec:Gaussianity_Measure}

We characterize the cluster environment through the gaussianity of the projected velocity distribution. Comparing and classifying distributions is a long-standing problem in statistics due to the difficulty on measuring the distance between two distributions (\citealt{HEL1904}, \citealt{SCH02}). For simplicity, we may assume that multimodal expression patterns results from multiple interacting groups; bimodality expresses two groups in interaction or a perturbation of a single Gaussian distribution; and unimodality represents a system close to virialization. The problem is then how to find multiple modes (Gaussians for instance) in a distribution. Thus, we can either identify a certain mixture of multiple modes (Gaussians) justifying the observed distribution or we determine how far from a Gaussian a distribution is. Each approach has its pros and cons. dC17 tackle the problem by creating realizations representing a perfect mixture of Gaussians and show that the later method (measuring the distance from a Gaussian) works better. This simplification, although not representing what happens in real clusters, serves as a guide to study multimodality modeling. 

In this work, we make use of the classification method presented in dC17, based on
the Hellinger Distance (HD), that measures the distance between two
discrete distributions, $\rm P_{1}$ and $\rm P_{2}$, and is expressed
as:
\begin{equation}
\label{eq:HD_equation}
\rm    HD^{2}(P_1, P_2)  = 2 \sum_{x} \left[\sqrt{p_{1}(x)} - \sqrt{p_{2}(x)}  \right]^{2},
\end{equation}
where $\rm p_{1}$ and $\rm p_{2}$ are the two probability density
functions (PDFs) and $x$ is a random variable
\citep[see][for more details]{le2012asymptotics}.

A detailed HD characterization and the threshold between Gaussian and NG velocity distribution as a function of members galaxies number is presented in dC17. In a nutshell, dC17 probe the parameter space defining a bimodal distribution and find which parameter mostly affects the distance between the two distributions. The HD proved to be robust in distributions with at least 20 members within $\rm R_{200}$, which translates to a mass cutoff. The relation between $\rm M_{200}$ and $\rm N_{200}$ (where $\rm N_{200}$ is the number of B galaxies inside $\rm R_{200}$) yields a lower mass threshold of $\rm 10^{14} M_{\odot}$ for our sample. We also restrict our sample to systems with at least 70\% reliability, measured with a bootstrap technique, on the gaussianity classification. This reduces the sample from 319 to 177 clusters (split into 143 G and 34 NG). The ratio of G and NG clusters in our sample (G/NG $\sim 81\%$) is in agreement with previous works (e.g. \citealt{2013MNRAS.434..784R}). Our final sample comprises 6,578 B galaxies (4,817 in G clusters and 1,661 in NG) and 2,205 F galaxies (907 in G and 1,298 in NG).

\subsection{Derived Stellar Population Parameters}
\label{subsec:Derived_Parameters}

We select age, stellar metallicity ([Z/H]) and stellar mass ($\rm
M_{stellar}$) from the dC17 galaxy Catalog to characterize the
stellar population content. The estimates are derived from full
spectral fitting using the {\sc STARLIGHT} code
\citep{2005MNRAS.358..363C}, that fits the input galaxy spectra with a
superposition of pre-defined single stellar populations (SSP). The age
and stellar metallicity are estimated from the weighted sum of the combination
of SSP parameters that gives the best fit, and are only derived
for spectra without any anomalies (see
dC17 for more details). The adopted stellar models
are based on the Medium resolution INT Library of Empirical Spectra 
\citep[MILES,][]{2006MNRAS.371..703S}, that features an almost constant
spectral resolution ($\rm \sim 2.5$\,\AA). The SSP basis grid has a constant
$\rm log(Age)$ steps of 0.2\,dex from 0.07 to 14.2 Gyr and includes SSPs
with [Z/H] = $\{-1.71, -0.71, -0.38, 0.00, +0.20\}$. $\rm M_{stellar}$ is
derived within the fiber and then extrapolated to the whole galaxy
by computing the difference between fiber and model magnitudes
in the $z$-band \citep{2012ApJ...752L..27T}, assuming no gradients in
the population content. Therefore,  $\rm M_{stellar}$ is given by:
\begin{equation}
\label{Mstellar_equation}
\rm    \log(M_{stellar}) = \log(M_{stellar})' + 0.4\,(m_{fiber,z} - m_{model,z}).
\end{equation}
Spectral fitting codes allow arbitrary weighting and there are two
commonly adopted methods in the literature: 1) luminosity-weighted
parameters trace mainly younger stellar population properties; while
2) mass-weighted parameters are more closely related to the cumulative
galaxy evolution \citep[see,][]{2020MNRAS.491.5406T}. In the following,
we use luminosity-weighted parameters, less prone to biases.  In this
case, age will be closely related to the last episode of star
formation.

\subsection{Assessing Derived Stellar Population Uncertainties}

The stellar population parameters have an intrinsic uncertainty as
any derived quantity. Part of this uncertainty comes from the observed
spectra, which can vary over different observations. To tackle this, 
we use a subsample of SDSS-DR7
galaxies with repeated observations. This set covers the same
redshift and magnitude range of our main sample and is limited
to observations with a signal-to-noise ratio greater than 20
in the $r$ band. This results in 6,148 observations of 2,543 galaxies. We calculate the
uncertainty separately for the two luminosity regimes, bright (B) and
faint (F). A direct comparison yields the residuals shown in
Table~\ref{table:starlight_errors}. Column (1) shows the luminosity regime
considered for the residuals; columns (2), (3) and (4) list the
residuals and the errors in Age, [Z/H] and $\rm M_{stellar}$,
respectively. As expected, the stellar population parameters of faint
galaxies have higher uncertainties than bright galaxies. From here on
we adopt the measurement errors shown in Table \ref{table:starlight_errors}. We compare these measurement errors with statistical uncertainties where it is relevant.

\begin{table}
\centering
\caption{Uncertainties in stellar population parameters.}
\label{table:starlight_errors}
\resizebox{0.85\columnwidth}{!}{\begin{minipage}{\columnwidth}
\begin{tabular}{c|ccc}
\hline
$\Delta$ & Age (Gyr)       & [Z/H] (dex)     & $\rm log(M_{stellar}/M_{\odot})$ (dex) \\ \hline
BG       & $0.18 \pm 0.35$ & $0.003 \pm 0.019$ & $0.01 \pm 0.01$                       \\
FG       & $0.26 \pm 0.61$ & $0.006 \pm 0.032$ & $0.06 \pm 0.03$                       \\ \hline
\end{tabular}

\end{minipage}}
\end{table}

\subsection{Additional Galaxy Properties}
\label{subsec:Additional_Information}

In addition to the stellar population parameters, we retrieve star
formation rates (SFR), morphology and color gradients, further tracers
of galaxy evolution. In this subsection, we describe relevant galaxy
properties retrieved from other works than dC17.

\subsubsection{SDSS-DR13 Spectroscopic Information}

We retrieve Star Formation Rates (SFR) from the MPA-JHU catalog, that
provides measurements for all SDSS-DR13 galaxies with reliable spectra
\citep{2017ApJS..233...25A}. The available SFRs were computed
following \cite{2004MNRAS.351.1151B}, that use the $\rm H_{\alpha}$
line luminosity measured within the spectroscopic fiber, correcting
the aperture effect with photometry. Our query of the SDSS-DR13
database results in SFR estimates for all galaxies, except for a set of 384
galaxies, which constitute a small percentage ($\rm \sim 2.46\%$) of
the total sample. We discard these galaxies only in the SFR-dependent
analysis.

\subsubsection{Morphological Characterization}

Galaxy morphology is intimately related to stellar population parameters
and thus galaxy evolution \citep[see, e.g.,][]{1994ARA&A..32..115R}. Here we
use the TType parameter as a tracer of galaxy morphology. It was
first introduced by \cite{1963ApJS....8...31D} to classify lenticular (S0) 
galaxies. Each galaxy is assigned a number based on morphological visual
classification. Elliptical-like morphologies are denoted by TType$<$0,
while TType$>$0 represents disk galaxies.

We select TType estimates from the \cite{2018MNRAS.476.3661D}
catalog, which uses deep learning algorithms based on  Convolutional
Neural Networks (CNN) to classify the morphology of 670,722 SDSS
galaxies. TType classifications are in the range $[-3,10]$. Similar
to SFR, this parameter is not available for all galaxies in our sample
and we discard non-classified galaxies only in the TType dependent
analysis. Statistically, this corresponds to a minor percentage of the
whole sample, $\sim 1.33\%$ (i.e 204 galaxies).

\subsubsection{Information from the Korea Institute for Advanced Study Value-Added Galaxy Catalog}

In our work, we add color gradient estimates from the Korea Institute
for Advanced Study Value-Added Galaxy Catalog
\citep[KIAS-VAGC,][]{2010JKAS...43..191C}. The KIAS-VAGC catalog
provides galaxy spectroscopic and morphological information for
$593,514$ galaxies from the SDSS-DR7 main galaxy catalog and
$10,497$ from other galaxy catalogs (see \citealt{2005ApJ...635L..29P}).
The color gradient is computed using
the color indices in the g and i bands. The $\rm g-i$ color gradient is
derived as:
\begin{equation}
\label{eq:grad_gi}
 \rm   \nabla (g-i) = (g-i)_{0.5R_{p}<r<R_{p}} - (g-i)_{r<0.5R_{p}}, 
\end{equation}
where $\rm (g-i)_{x}$ denotes the g-i index where the condition x is
satisfied, and $\rm R_{p}$ is the Petrosian Radius in the
i-band. Equation~\ref{eq:grad_gi} implies that more negative values
correspond to bluer colors in the galaxy outskirts.

We use a 1.5\,arcsec threshold in a positional cross-match between the
KIAS-VAGC and our sample to select the appropriate estimates. The
upper limit is defined empirically. We do not find reliable estimates
for 304 galaxies, representing $1.96\%$ of our sample. As in the
previous cases, we discard non-classified galaxies only in the 
analysis pertaining to color gradients.

\section{Structure and Composition of G and NG Clusters}
\label{sec:Structure}
\begin{figure}
    \centering
    \includegraphics[width = 1\columnwidth]{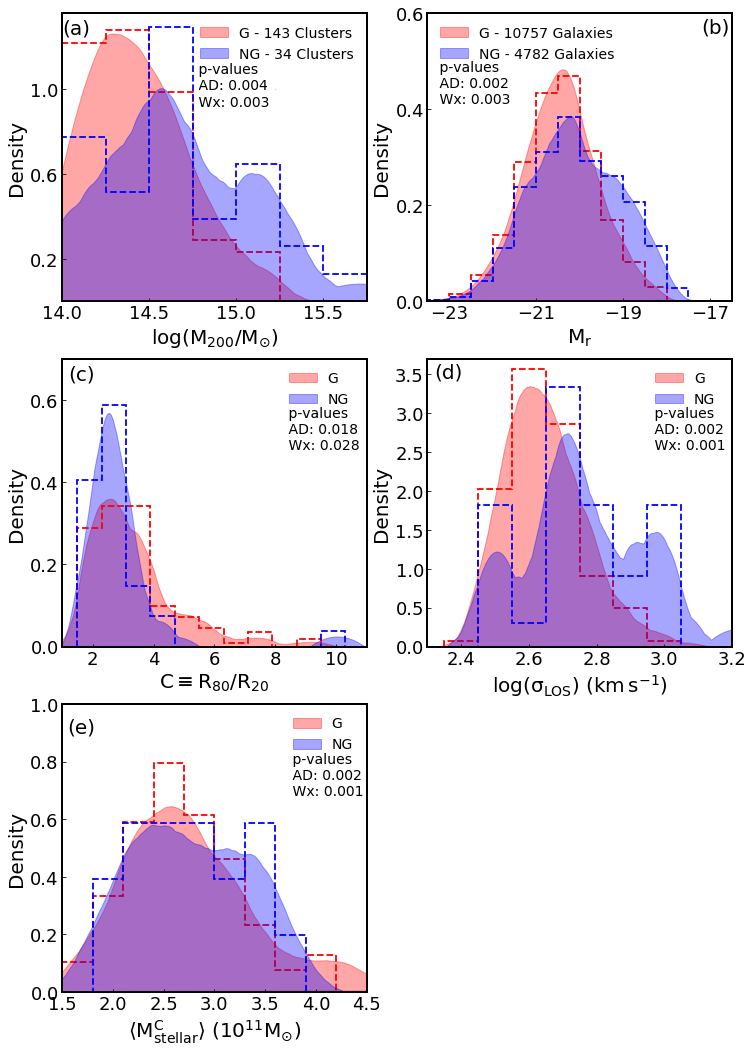}
    \caption{Comparison of the $\rm M_{200}$ (top-left), $\rm M_{r}$
      (top-right), $\rm C \equiv R_{80}/R_{20}$ (center-left), $\rm log(\sigma_{LOS})$
      (center-right) and cumulative stellar mass ($\rm M_{stellar}^{C}$) distribution according to the gaussianity
      classification (G or NG cluster). In each plot we show the
      resulting p-value of the Anderson-Darling (AD) and Wilcoxon (Wx)
      statistical tests. NG clusters are more massive, less
      concentrated (see text for the concentration measurement adopted here),
      larger virial radius and have an excess of fainter
      galaxies with respect to G systems. }
    \label{fig:concentration}
\end{figure}

In this work, the difference between G and NG clusters plays a major
role. In a first step, we focus our attention on characterizing the
structure and distribution of galaxy member properties in each class.
In Fig.~\ref{fig:concentration} we show the
distributions of: a) logarithmic virial mass ($\rm log(M_{200}/M_{\odot})$);
b) r-band absolute magnitude; c) a proxy for the concentration of
stellar mass in each cluster, defined as $\rm R_{80}/R_{20}$, where
$\rm R_{x}$ is the projected radius within which the stellar
mass represents x\% of the total stellar mass within $\rm R_{200}$; d) velocity dispersion along the line-of-sight; and e) the distribution of cluster mean stellar mass of bright galaxies within $\rm R_{200}$ ($\rm \langle M_{stellar}^{C} \rangle$)\footnote{We consider only bright galaxies to avoid bias due to faint galaxies in clusters with z>0.04} We
compare the distributions using two different statistical tests: 
Anderson-Darling (AD) and Wilcoxon Rank Test (Wlx) (see
\citealt{engmann2011comparing} and \citealt{gehan1965generalized} for
a review of both)\footnote{The adopted significance level threshold is
$\alpha = 0.05$}. The results are shown in each panel. Aside from the histograms, we add a kernel smoothed curve, shown as shaded area, which is derived directly from the dataset using an Epanechnikov Kernel Density estimator \citep{1986desd.book.....S} with a bandwidth equal to 1.5 times the bin size. We find that G and NG clusters have statistically different distributions. In panel (a), we note that NG clusters tend to have
higher values of $\rm M_{200}$ in comparison to G clusters. Namely, we find
that 32.4\% (11/34) of NG clusters have $\rm log(M_{200}/M_{\odot})>14.75$, while
this fraction decreases to 6.3\% (9/143) in G clusters. The majority
of G systems ($\sim 71.3\%$) have $\rm log(M_{200}/M_{\odot})<14.5$. Panel (b)
shows an excess of fainter galaxies\footnote{Not to be confused with
 the faint regime defined in Section~\ref{subsec:Data_Select}} in NG
clusters in comparison to G systems. We find that 30.5\% of NG
cluster members have $\rm M_{r} \geq -19.5$, while the equivalent cut
in absolute magnitude yields 14.1\% of galaxies in G systems.
Panel (c) shows the distribution of concentration in G and NG clusters,
revealing that NG cluster galaxies are less concentrated than their
G cluster counterparts. Only one NG cluster reaches C$>$4.7. In panel (d), we observe an excess of NG clusters with higher velocity dispersion in comparison to G clusters. NG clusters are presumed to be found in a non-virialized state, so that the expected velocity dispersions are higher, and, possibly, the estimated mass may also be an overestimate of the real cluster mass. In any case, this is one more piece of evidence regarding the different state of NG clusters with respect to G systems. Finally, we note in panel (f) an excess of NG clusters with mean stellar mass $\rm 3 \leq \langle M_{stellar}^{C} \rangle < 4 \times 10^{11} M_{\odot}$. In other words, we find that NG clusters have higher virial mass and radii, an excess of fainter galaxies, are less concentrated, contain more massive B galaxies and have higher velocity dispersion in comparison to G systems. 

We give special attention to the mass mismatch of G and NG clusters. The excess of massive NG clusters is very likely related to the recent merger history of such systems. \cite{2019MNRAS.490..773R} use simulations to investigate the dynamical history of G and NG clusters and find that NG clusters suffered their last major merger more recently. Our results suggest that the difference in mass between G and NG likely follow from recent accretion. Additionally, \cite{2012MNRAS.424..232W} show that galaxy properties depend on halo mass, which can also be related to the separation between G and NG clusters. An usual approach to avoid mass bias is to compare mass-matched samples. However, we lack G clusters in the faint regime with $\rm log(M_{200}/M_{\odot})>14.6$, which prevents the analysis of such a mass-matched sample in this regime. Namely, a mass-matched sample in the faint regime would contain only 5 clusters. Nevertheless, we build a mass-matched sample in the bright regime and compare the results with the unmatched ones. Qualitatively, we find that our results do not depend on the use of a mass-matched sample. We present in Section \ref{sec:relations} how the mass mismatch affects our results quantitatively.

\begin{figure*}
    \centering
    \includegraphics[width = 0.85\textwidth]{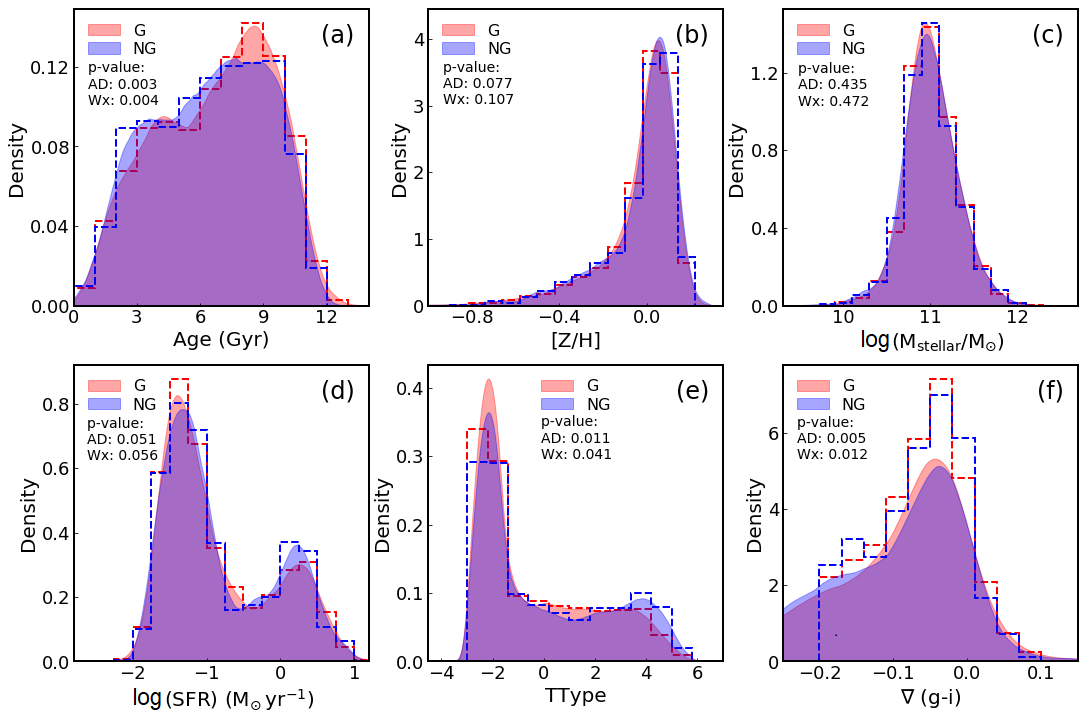}
    \caption{Distributions of age (a), [Z/H] (b), $\rm log(M_{\rm stellar}$) (c), log(SFR) (d), TType (e) and $\nabla$(g-i) (f) for bright galaxies in G and NG clusters. In each panel we also show the resulting p-values of an AD and Wx statistical test. The distributions of Age and TType of BG are statistically different between G and NG clusters.}
    \label{fig:bright_hist}
\end{figure*}

\begin{figure*}
    \centering
    \includegraphics[width = 0.85\textwidth]{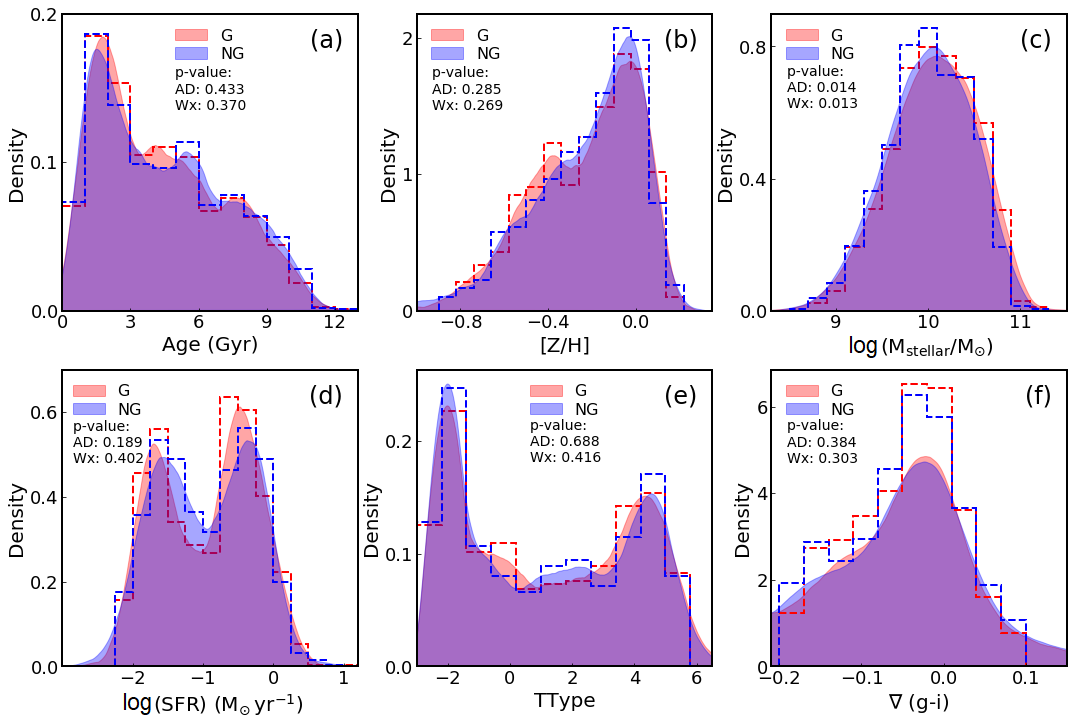}
    \caption{The same as Figure \ref{fig:bright_hist}, but for F galaxies. Differently from the bright regime case, only $\rm \log(M_{\rm stellar})$ have statistically different distributions.}
    \label{fig:faint_hist}
\end{figure*}

Regarding galaxy properties, we compare the distribution of age,
[Z/H], $\rm M_{stellar}$, SFR, TType and color gradient ($\rm \nabla (g-i)$)
of G and NG cluster members in Figs.~\ref{fig:bright_hist}
(B galaxies) and \ref{fig:faint_hist} (F galaxies). In the bright
case, we notice that G and NG clusters have statistically different
distributions of age, TType and $\rm \nabla (g-i)$. In panel (a) of
Fig.~\ref{fig:bright_hist}, we notice in G clusters an excess of B
galaxies with $\rm age > 7.5$ Gyr in comparison to NG systems. Regarding
galaxies with $\rm age < 7.5$ Gyr, the excess is seen in NG systems. In
panels (b), (c) and (d), we note that G and NG clusters have a similar
distribution of [Z/H], $\rm M_{stellar}$ and
SFR. In panel (e), NG clusters feature an excess of B
galaxies with $\rm TType > 3$, which translates to an excess of B galaxies
with disk-like morphology. Finally, although the color gradient distributions (panel f) do not show great differences visually, statistical tests indicate significant differences between G and NG clusters. Namely, we find a slight excess of galaxies with steeper color gradients in NG clusters. Extending the same analysis to F galaxies, shown in Fig. \ref{fig:faint_hist}, we find that $\rm M_{stellar}$ is the only parameter to show statistically different distributions. We note a slight excess of faint galaxies with $\rm log(M_{stellar}/M_{\odot}) > 10.2$ in G clusters in comparison to NG clusters. In all other panels we notice that the p-value analysis is unable to reject the null hypothesis that both distributions represent subsamples of the same parent distribution.

Following previous works \citep[dC17]{2017MNRAS.467.3268R,2019MNRAS.487L..86D}, we would expect differences between G and NG cluster members mainly in the faint regime. However, the comparison of global distributions of F galaxy properties shows small (if any) differences. Furthermore, we find a different systematic in comparison to dC17, which may result from the use of the distribution itself instead of the cumulative distributions. Therefore, we use the PPS to better disentangle the expected differences.

\section{The Projected Phase Space}
\label{sec:thePPS}

In this section, we describe how the PPS is defined in G and NG
clusters, and contrast the results. The PPS diagram consists of
two projected dynamical quantities: the peculiar velocity --
measured along the line of sight ($\rm V_{LOS}$) -- and the cluster-centric
distance -- projected on the observer's plane ($\rm R_{LOS}$). We discard sign differences in the velocities (approaching/receding) and take the absolute value of $\rm V_{LOS}$, following standard practice due to the existing symmetry along the line of sight (e.g. \citealt{2011MNRAS.416.2882M,2017ApJ...843..128R,2019MNRAS.484.1702P,2020ApJS..247...45R}). The radial distance is given in units of $\rm R_{200}$ and the velocity axis is normalized by the velocity dispersion of the cluster along the line-of-sight (simply $\sigma$, hereafter) in order to enable comparisons between different systems. The normalization is based on a homology hypothesis, that states that systems are structurally equal, aside from a characteristic radius and velocity dispersion.

A major concern in the PPS construction is that clusters in our sample
have a large range of richness, ranging from 25 to 900 galaxies. To
avoid biases due to the different number of members, we stack
all the galaxies belonging to the same cluster class and luminosity
regime into a single PPS. Hence, we end up with four stacked PPSs: two 
for G clusters (B and F) and two for NG clusters (B and F).

\subsection{Discretizing the Projected Phase Space}

The PPS approach enables the identification of cluster members according
to their infall time. Although in the full 6D phase space galaxies have
well-defined trajectories as they enter the respective  cluster 
(e.g. \citealt{2010MNRAS.408.2442W}), in the PPS this trajectory is
degenerate due to projection effects. Direct 
tracing of galaxies at different times since infall is not possible with 
observational data. Instead, cosmological simulations are used to define
different regions in the PPS that correspond to the loci of galaxies at different
times since infall. We make use of two different ways of slicing
the PPS: 1) following R17, that use the YZiCS
simulation to study the numeric density of galaxies at different
times since infall occupying specific regions in the PPS. We refer to
these regions as ``Rhee Regions'' hereafter. They provide a
probabilistic approach to each region in the PPS; and 2) the
separation presented in P19, also based on the YZiCS simulation, but
defining analytic quadratic functions to fit the observed distributions 
of time since infall in the simulated stacked PPS. The PPS is then segmented in regions constraining galaxies to a narrow range of infall time, as done by P19, where the mean and the variance of each slice are listed in their Table 1. These PNZs (Pasquali New Zones, hereafter) are defined in decreasing mean infall time from PNZ 1 (innermost region, $\rm t_{inf} \sim 5.4$\,Gyr) to PNZ 8 ($\rm t_{inf} \sim 1.4$\,Gyr). Variances in the mean infall time of each PNZ range from $\rm \sim 1.5 - 2.5$ Gyr.

Although R17 and P19 use the same cosmological simulation (YZiCS), the way they slice the PPS is markedly different. The main differences are: 1) while R17 slice the PPS in 5 regions, P19 do it in 8, 2) P19 do no include interlopers into their definition, while Rhee Regions take them into account (see Fig.~6 in R17); and 3) P19 limit their analysis to $\rm R_{200}$, while R17 extend to $\rm 2R_{200}$. We then compare how our results depend on the way we slice the PPS. We find the same trends between G and NG either using P19 or R17 approaches. Appendix A presents a more detailed discussion regarding this comparison. In this work, we use the P19 method of slicing the PPS due to its direct relation with infall time.

A word of caution is needed regarding the way the PPS is discretized, as in R17 and P19. We may ask which role is played by backsplash galaxies when we examine the PPS since they may suffer a partial quenching and have their stellar population properties modified when confronted to other galaxies in the cluster. Even using cosmological simulations it is not easy to define where these systems dominate in the PPS. For instance, \cite{2011MNRAS.416.2882M} find that the locus of their dominance is around $\rm 1.0 \leq R_{LOS}/R_{200} \leq 1.5$ $\times$ $\rm 0 \leq |V_{LOS}|/\sigma \leq 1.0$, despite they represent only $\sim 30\%$ of the galaxies in this box. To address the question of how backsplash galaxies can affect the results here obtained, we compare the distributions of backsplash galaxy properties in G and NG clusters similarly to Fig.~\ref{fig:bright_hist}. An important caveat is that in this case the distributions contain roughly $\sim 10$ times less points (galaxies) than the distributions shown in Figs.~\ref{fig:bright_hist} and \ref{fig:faint_hist}. Nevertheless, we find that backsplash galaxy properties have similar distributions compared to those shown above, especially in the faint regime. The similarity indicates that backsplash galaxies are affected by the environment just as any other set of galaxies. Therefore, comparison between galaxies within $\rm R_{200}$ and the backsplash-limited ones (outside $\rm R_{200}$) suggest that they do not influence our results significantly.

\subsection{Galaxy Distribution in the PPS}
\label{sec:PPS_distribution}

In a first approximation, differences in cluster galaxy population may
translate to a different distribution in the PPS. In this section, we
explore how galaxies are distributed in the PPS of G and NG
clusters. Table~\ref{table:galaxies_distribution} shows the number of
galaxies in each PNZ for G and NG systems, for both B and F
regimes. Column (1) lists the cluster class; column (2) lists the
luminosity regime; columns (3) to (10) show the number of galaxies in
PNZs 1 to 8, respectively; and column (11) lists how many galaxies are located 
beyond $\rm R_{200}$. 41\% of the B galaxies in NG clusters are beyond
$\rm R_{200}$, while in G clusters this percentage decreases to
28\%. F galaxies also show a similar trend: 28\% and 34\% of the F
galaxies are beyond the $\rm R_{200}$ of G and NG clusters,
respectively. These numbers unequivocally confirm that NG clusters show
an excess of galaxies beyond $\rm R_{200}$.

\begin{figure}
    \centering
    \includegraphics[width = \columnwidth]{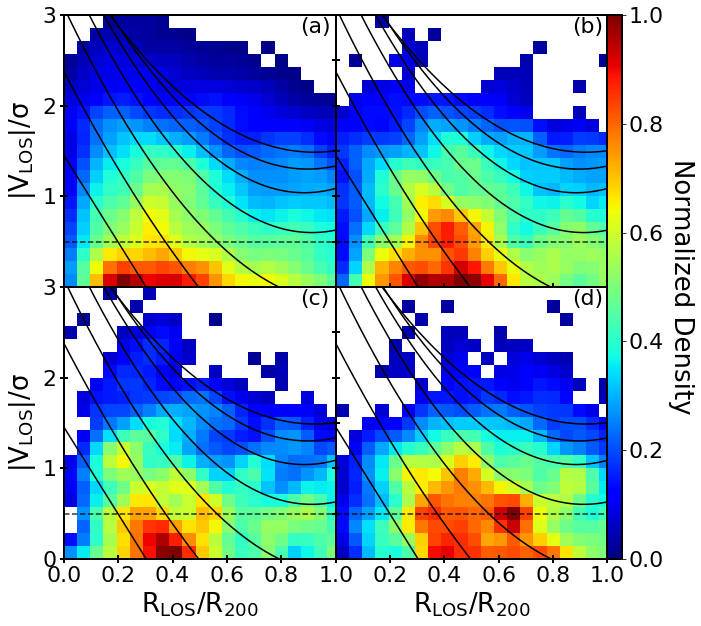}
    \caption{Normalized density distribution of galaxies in the PPS of
      G clusters (left) and NG clusters (right). We separate galaxies
      according to the two adopted luminosity regimes: bright on top panels and
      faint in the bottom panels. We note that for NG clusters, higher
      density regions extend to higher velocities in comparison to G
      clusters. We highlight this trend by including a $\rm
      |V_{LOS}|/\sigma = 0.5$ dashed line in the four panels.}
    \label{fig:PPS_Density}
\end{figure}

In Fig.~\ref{fig:PPS_Density} we show the normalized density of B
and F galaxies in the PPS of G and NG clusters. We limit the PPS to
$\rm R_{200}$, where the PNZs are defined and enable a connection between
infall time and PPS location. We divide the PPS into bins of
$\rm 0.15 |V_{LOS}|/\sigma \times 0.05 R_{LOS}/R_{200}$ to guarantee
a good sampling of the PNZs, adapting the bins to the aspect ratio of the
diagram (note the ordinate extends a factor of 3 with respect to the
abscissa, in normalized units). This slicing results in 12 and 6 B galaxies per bin (on average) in G and NG clusters, respectively. In the faint regime, we find lower values ($\sim 5$ galaxies per bin in both cases). We use these bins to map the distribution of galaxies in PPS, 
performing a convolution in the resulting matrix (defined by the
number of galaxies per bin)
using a Gaussian kernel with $\rm FWHM = 1.88$ ($\rm \sigma_{std} = 1$)\footnote{The FWHM/standard deviation was empirically defined, but does not affect the resulting trends.}. The kernel size is defined to slightly reduce the noise and preserve the global trends. Bins without galaxies are carefully treated, applying an interpolation in ``empty'' bins that have at least 50\% of their neighbouring bins occupied. Note that by applying this interpolation scheme, we assume that transitions within these bins in PPS space are sufficiently smooth. 

\begin{table}
\caption{Number of galaxies in each PNZ.}
\label{table:galaxies_distribution}
\resizebox{0.83\columnwidth}{!}{\begin{minipage}{\columnwidth}

\begin{tabular}{c|c|ccccccccc}
\hline
                    & PNZ & 1   & 2   & 3   & 4   & 5   & 6   & 7  & 8   & $\rm R>R_{200}$ \\ \hline
\multirow{2}{*}{G}  & BG  & 392 & 611 & 723 & 824 & 391 & 176 & 93 & 255 & 1352            \\
                    & FG  & 58  & 125 & 249 & 150 & 66  & 33  & 19 & 41  & 289             \\ \hline
\multirow{2}{*}{NG} & BG  & 101 & 157 & 126 & 229 & 141 & 66  & 26 & 69  & 623             \\
                    & FG  & 47  & 128 & 182 & 249 & 119 & 52  & 22 & 58  & 441             \\ \hline
\end{tabular}

\end{minipage}}
\end{table}

We note in Fig.~\ref{fig:PPS_Density} that indistinctly for B or F
galaxies and G or NG clusters there is an offset of approximately
$\rm R_{LOS} \sim 0.4 R_{200}$ between the cluster center and the higher
density envelope (redder colors). This offset may be due to
projection effects or a limitation when defining the cluster centers 
with galaxies, that represent a minor fraction of the
cluster mass, compared to the gas and dark matter components. This
issue lies beyond the scope of this paper but will be further
investigated in future work (Sampaio et al. in prep). Despite this
offset, the galaxy distributions are in agreement with both
observational and simulated data from \cite{2020ApJS..247...45R},
guaranteeing that this offset does not represent a sample bias.

Comparison between galaxy distributions of G and NG clusters in the PPS 
shows that the higher density envelope (redder colors) in NG
clusters extends to higher velocities than in the G case, for both B
and F (see the dashed line in Fig.~\ref{fig:PPS_Density}). We give special attention to PNZ 8, where we expect to find first infalling galaxies. However, it is important to consider that, in both B and F regimes, G and NG clusters have a different number of galaxies (see Section~\ref{subsec:Gaussianity_Measure}). To detail differences in PNZ 8 we then follow: 1) from the cluster class with more galaxies (G in B and NG in F) we randomly select N galaxies, where N is the number of galaxies in the less rich class (NG in B and G in F); 2) calculate the ratio of galaxies in the PNZ 8 between the equal-size samples; and 3) we repeat this procedure 1,000 times. This guarantees that differences of G and NG cluster PNZ 8 are not due to one sample being larger. A comparison shows that G clusters have an excess of $\sim 27\%$ of B galaxies in the PNZ 8 with respect to NG clusters. In the faint regime, we find that PNZ 8 is roughly equally occupied in both cases. Namely, we find a mean ratio of $0.99$.  

Regarding the excess of bright galaxies in the PNZ 8 of G systems with respect to NG ones, no firm conclusion can be drawn. However, we should stress that dC17 (see their Figure 7) find a higher kurtosis for the bright galaxies in G systems with respect to the NG ones. Datasets with high kurtosis tend to have heavy tails while datasets with low kurtosis tend to have light tails. Although this is not an explanation, it shows consistency. Furthermore, normally we would expect a more populated “virialized core”, i.e. low PNZ number in G clusters. What we find is that NG clusters have fewer galaxies in high PNZ regions, i.e. mainly at high $\rm V_{LOS}/\sigma$, so this would mean that the distributions of NG clusters are sub-gaussian, or platykurtic. The reason as to why the kurtosis is negative needs further investigation. The point above justifies that NG systems have higher $\sigma$ (as an equivalent width of the velocity distribution), but not necessarily that the tails of the distribution are filled with respect to Gaussians, i.e. not enough high-velocity galaxies as they are still in a preliminary infall state. We plan on investigating this particular issue using cosmological simulations like Illustris and YZiCS in a future work.

\section{Exploring various galaxy properties in Projected Phase Space}
\label{sec:locus_properties}
Environmental quenching drastically affects galaxy evolution. Previous
works show that the quenched fraction of galaxies within clusters is a
function of clustercentric radius \citep[e.g.,][]{2020ApJ...899...85S}. However, RPS and TML (see Section~1) are also conditional on the incoming velocity of the infalling galaxy (see R19
for an example). The PPS provides a powerful tool to understand how galaxy properties are affected in high-density environments, taking into account both velocity and position. In this section, we study the distribution on the PPS of median age, [Z/H], SFR, TType, $\rm \nabla (g-i)$, $\rm M_{stellar}$ in G and NG clusters. We use a similar approach as in Fig.~\ref{fig:PPS_Density} to study the corresponding distributions. Furthermore, we calculate the variance in each pixel for each parameter to compare with measurement errors. In general, the variance in the pixel is greater than the measurement errors. For instance, we find that the map of the B galaxies Age has a mean variance of 0.74 Gyr. In the following, we describe differences greater than the mean variance.

\begin{figure*}
    \centering
    \includegraphics[width = \textwidth]{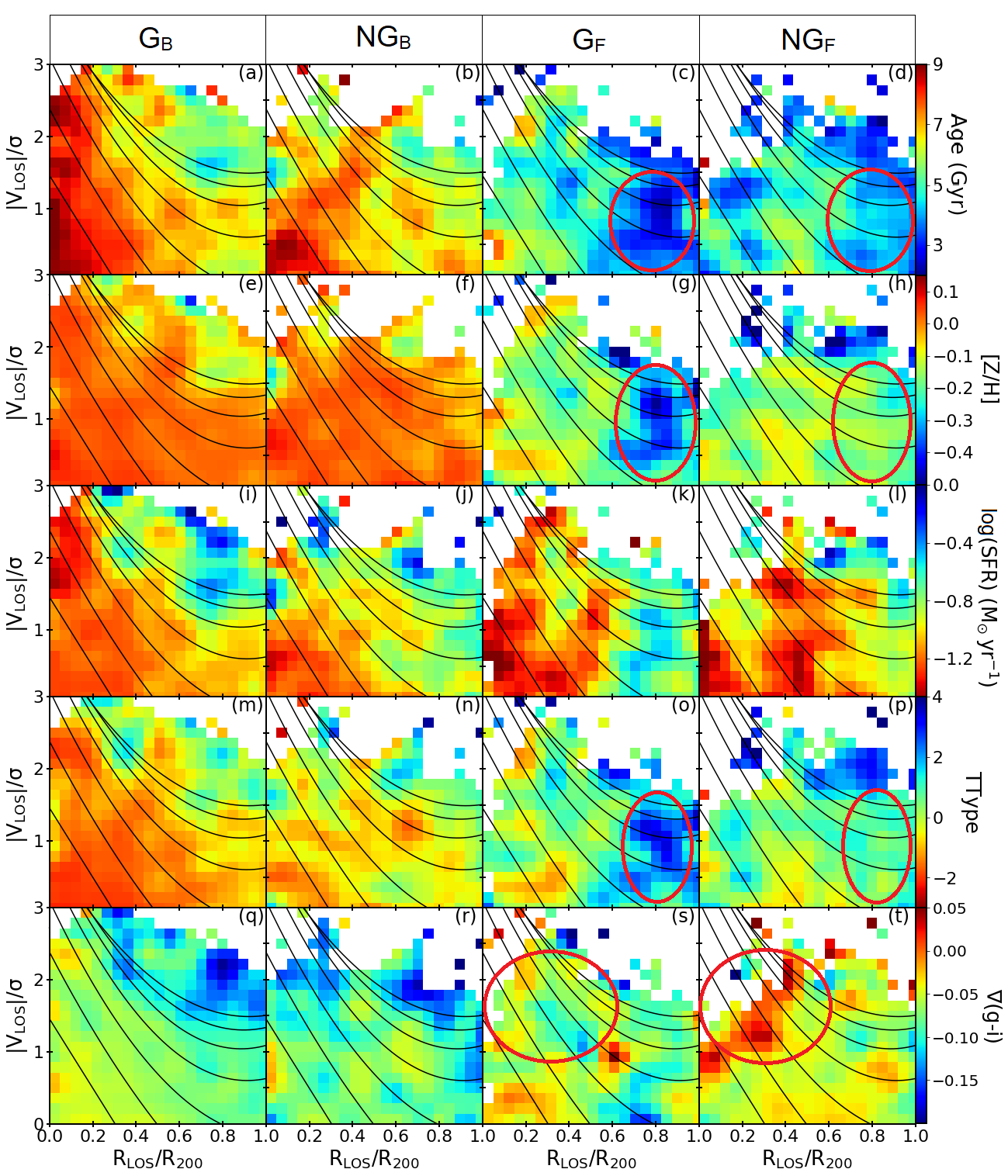}
    \caption{From top to bottom we present the distribution of age,
      [Z/H], $\rm \log(SFR)$, TType and $\nabla$(g-i), respectively, over
      the PPS of G and NG clusters and luminosity regime (B or F). The red ellipses highlight differences between galaxy properties in G and NG clusters, in the faint regime.}
    \label{fig:full_grid}
\end{figure*}

Fig.~\ref{fig:full_grid} displays (from top to bottom) the
distribution on the PPS of median stellar age, [Z/H], $\rm \log(SFR)$, TType,
and $\rm \nabla(g-i)$ for B (leftmost two columns) and F (rightmost
two columns) galaxies, in G and NG clusters, as labeled. The regions delimiting the PNZs are shown as solid lines and classify galaxies according to their infall times. We tested our results against statistical noise in the following way: 1) we randomly divide our sample (177) into two halves; 2) we perform the same analysis as in Fig.~\ref{fig:full_grid}; and 3) compare the differences between the two half-samples. When clusters are randomly chosen we do not find significant differences. This ensures that the differences between G and NG clusters are not due to statistical noise. We also use this test as guidance to define the significant differences between G and NG clusters.

Comparison of B galaxy property distributions in G and NG clusters (two first columns of Fig.~\ref{fig:full_grid}) shows some general trends. In summary, we find: 1) NG clusters have more mixed distributions, whereas G clusters have smoother distributions concerning the PNZs; 2) we find that both [Z/H] and color gradient show an almost uniform distribution over the PPS, indistinctly for G and NG clusters; and 3) B galaxies with high $\rm |V_{LOS}|/\sigma$ in $\rm PNZs < 4$ of G clusters (see Section \ref{sec:PPS_distribution}) are older, less star-forming and more elliptical than its counterpart in NG cluster. More quantitatively, a comparison of panels (a) and (b) shows that the (inner) $\rm PNZs \leq 4$ of NG cluster have galaxies with $\rm age < 7$ Gyr, while the equivalent in G clusters is dominated by older galaxies, with $\rm age > 7$ Gyr. Namely, 70\% of the $\rm G_B$ galaxies in $\rm PNZ \leq 4$ regions have $\rm age > 7$ Gyr, whereas for $\rm NG_B$ clusters this percentage is 53\%. In panels (e) and (f), the mean [Z/H] of $\rm G_B$ and $\rm NG_B$ galaxies is $-0.020 \pm 0.001$\footnote{Hereafter the shown values are calculated using the galaxy distribution itself, instead of the pixellated version of the PPS. However, comparison of the two methods shows differences lower than 5\% in both mean value and error.} and $-0.010 \pm 0.002$, respectively. Regarding SFR, both G and NG clusters show a trend of lower SFR with higher infall time (i.e. decreasing PNZ). The quenching is defined by a decrease in the SFR and this result highlights the cumulative effect of environmental quenching with respect to the time since infall. Despite this overall behaviour, a comparison of panels (i) and (j) shows that, $\rm G_B$ galaxies (panel i) with $\rm log(SFR) < -0.6$ occupy mainly PNZs 1 to 4. In contrast, the equivalent cut in $\rm NG_B$ galaxies (panel j) restricts the PPS to PNZs 1 to 3. In the PNZ 4 of NG clusters, we note an excess of galaxies with higher SFR in comparison to their counterparts in G systems. Examining the B galaxies morphological, TType distribution (panels m and n) we find similarities with the distributions of SFR, [Z/H] and Age, especially in the G case. This provides further evidence that the quenching of star formation correlates with morphological transformation. TType differences between G and NG clusters are restricted mainly to $\rm PNZs \leq 4$. We find that $\rm G_B$ galaxies in the $\rm PNZ \leq 4$ region have an average TType value of $-1.2 \pm 0.2$, substantially lower than the equivalent average in $\rm NG_B$ galaxies ($-0.5 \pm  0.2$). In the $\rm PNZ > 4$ areas, the mean difference between G and NG decreases to $\rm \Delta\langle TType \rangle (G-NG) \sim 0.2$ ($\rm\langle TType_{PNZ > 4}\rangle = $ -0.1 and 0.1 for G and NG clusters, respectively). Finally, the distribution of color gradient is similar in G and NG clusters (panels q and f). Galaxies with $\rm \nabla (g-i) < -0.1$ are found mainly in PNZ 8 and at $\rm |V_{LOS}|/\sigma > 1.7$. Lower values of color gradient mean bluer outskirts (see eq.~3). This trend suggests that B galaxies enter clusters with bluer colors on the outside and the difference decreases after traversing the PPS to PNZ 6.

The PPS analysis shows that, further to the differences found in the sample of B galaxies, we note more striking ones in the faint regime (two last columns of Fig.~\ref{fig:full_grid}). We highlight the main differences between F galaxies in G and NG clusters with red ellipses in Fig.~$\ref{fig:full_grid}$. Our results suggest that F galaxies (hence less massive) are mostly quenched by environmental effects, in agreement with works showing the stellar mass dependence on galaxy evolution \citep{2010ApJ...721..193P,2012MNRAS.424..232W}. In panels (c) and (d), we note that $\rm NG_F$ galaxies in the highlighted region are on average $1.4 \pm 0.4$\,Gyr younger than in the $\rm G_F$ case. On the other hand, $\rm NG_F$ galaxies beyond $\rm 0.6 \, R_{200}$ are slightly older by, on average, $0.9 \pm 0.2$\,Gyr concerning to $\rm G_F$ galaxies. Panels (g) and (h) show significant differences in average metallicity between G and NG clusters. $\rm G_F$ galaxies in the red highlighted region are more-metal-poor by, on average, $0.28 \pm 0.04$\,dex with respect to $\rm NG_F$ galaxies. However, within $\rm 0.6R_{200}$ there are no significant differences between G and NG clusters. More quantitatively, we find that $\rm G_F$ galaxies within $\rm 0.6 \, R_{200}$ are on average $0.010 \pm 0.002$\,dex more metal-rich than the $\rm NG_F$ counterparts, comparable to the uncertainty shown in Table~\ref{table:starlight_errors}. These results are in agreement with dC17, which further show that differences in the outskirt extend to $\rm 2 \, R_{200}$. Regarding F galaxies SFR, a trend of a core with quenched galaxies with a tail to higher velocities is visible in panels (k) and (l). In panel (k), we note a tail in the distribution with $\rm log(SFR) < -1.0$ that extends up to $\rm \sim 1.7 |V_{LOS}|/\sigma$. In NG clusters (panel k) this tail is more conspicuous. NG clusters also appear more mixed. Namely, we note a trail of quenched galaxies in $\rm 0.5R_{200} < R_{LOS} < 0.7R_{200}$ that extends to $\rm \sim 1.7 |V_{LOS}|/\sigma$ and that only PNZ 1 is fully occupied by galaxies with $\rm log(SFR) < -1.2$. As in the B regime, we note relevant similarities between the SFR, [Z/H], Age and TType distributions in both G and NG clusters, which further support the relation between morphology and stellar population properties. Regarding G vs NG, panels (o) and (p) show that $\rm G_F$ within the red ellipse in G clusters have higher TTypes compared to the same subset in NG clusters. Namely, we find an average TType value of $2.0 \pm 0.4$ and $0.9 \pm 0.4$ for G and NG clusters, respectively. At last, examining panels (s) and (t) we find that NG clusters have an excess of F galaxies with $\rm \nabla (g-i) > 0$ . This subset of
galaxies (highlighted by the red ellipse) is found in every PNZ, which may indicate that these F galaxies entered the NG cluster already with shallow/positive values of color
gradients.

\begin{figure}
    \centering
    \includegraphics[width = \columnwidth]{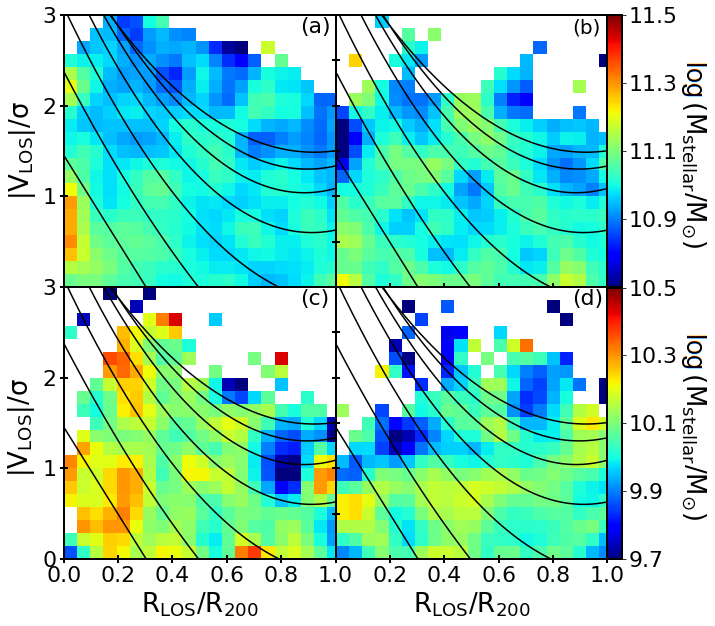}
    \caption{Panels (a) and (b): B galaxies $\rm M_{\rm stellar}$ distribution over the PPS of G and NG clusters, respectively. Panels (c) and (d): the same, but for F galaxies.}
    \label{fig:grid_mstellar}
\end{figure}

In Fig.~\ref{fig:grid_mstellar} we present the median $\rm M_{stellar}$
distribution in the PPS of G and NG clusters, with
galaxies divided into B and F. We separate this specific parameter
from the last 5 discussed since in this case the B and F split translates,
by construction, into different ranges in stellar mass. Therefore, we
select suitable ranges for each luminosity regime, while guaranteeing
that both have the same relative difference in $\rm log(M_{\rm stellar}/M_{\odot}$).
The plotting ranges are defined from the distribution of $\rm log(M_{stellar}/M_{\odot})$, as 
shown in Figs.~\ref{fig:bright_hist} and \ref{fig:faint_hist}.
Comparing panels (a) and (b), we see that the
distribution of $\rm log(M_{stellar}/M_{\odot})$ in G and NG clusters
for B galaxies are quite similar. We observe that in both G and NG
clusters, only $\sim 34.5$\% of B galaxies have $\rm
log(M_{stellar}/M_{\odot}) > 11.1$. In the faint regime, we note that
approximately 48\% of the F galaxies have $\rm
log(M_{stellar}/M_{\odot}) > 10.1$, corresponding to the upper part of
the defined range. In Section~\ref{sec:Discussion}, we discuss this
difference between B and F galaxies in the context of the downsizing
model \citep{2006MNRAS.372..933N}. In the faint regime, we also note
significant differences between G and NG clusters. Panel (c) shows
that within $\rm 0.3 \, R_{200}$ there is an excess of F galaxies with $\rm
log(M_{stellar}/M_{\odot}) > 10.2$ in G clusters. Additionally, in
panel (d) we note within $\rm 0.4 \, R_{200}$ NG clusters feature an excess of F
galaxies with $\rm log(M_{stellar}/M_{\odot}) < 10.0$.

We find that galaxy property distributions in G and NG clusters are significantly different from the PPS point of view. Despite global similarities (see Section \ref{sec:Structure}), the PPS analysis shows they are differently distributed in such systems. Furthermore, we find good agreement between galaxy properties variation in the PPS and PNZs (and infall time, consequently). This suggests a more straightforward relation between galaxy properties and infall time as a consequence of the cumulative effect of environmental quenching.

\section{The Relation Between Galaxy Properties and Infall Time in G and NG Clusters}
\label{sec:relations}
\begin{figure*}
    \centering
    \includegraphics[width = \textwidth]{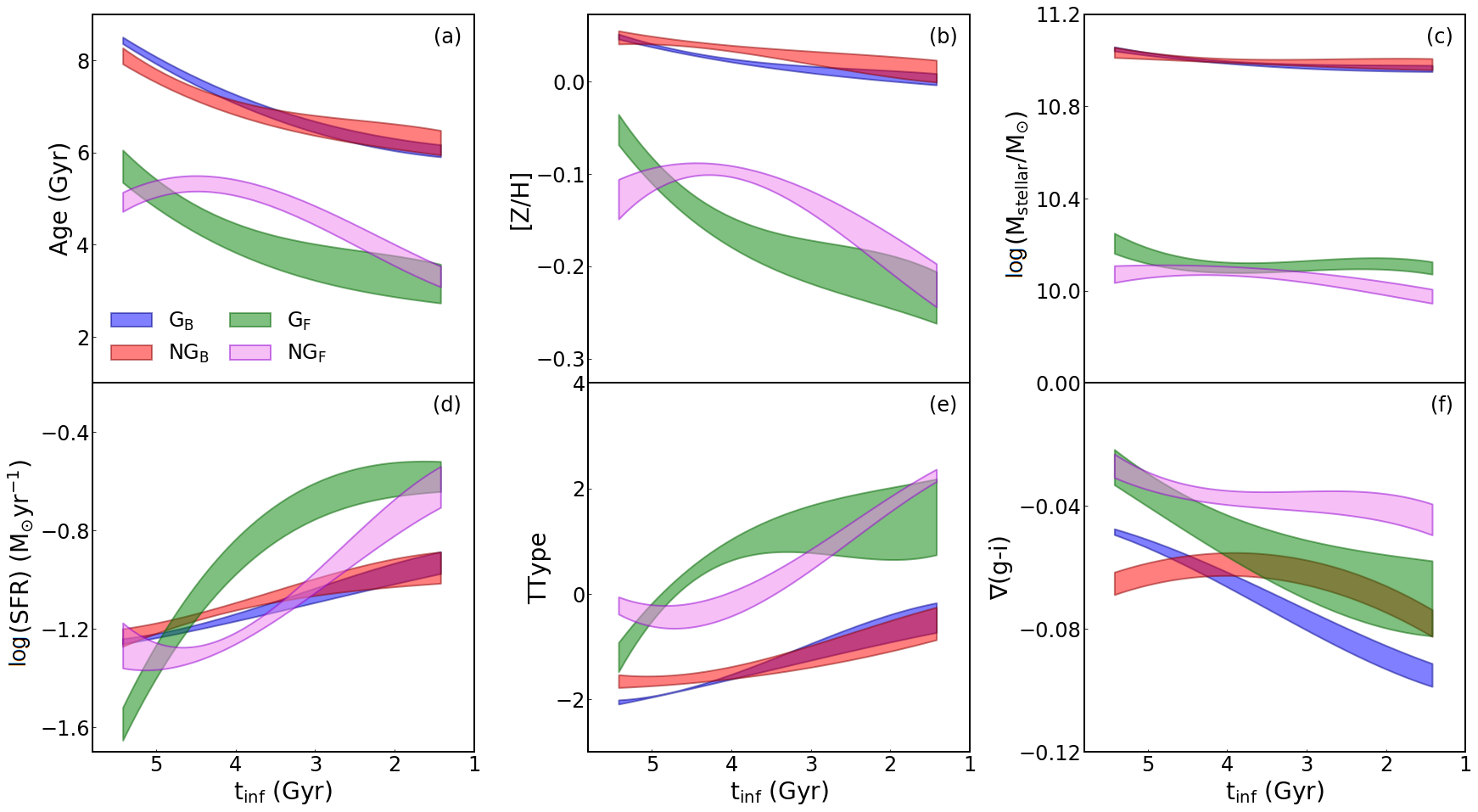}
    \caption{From panel (a) to (f): Relation between time since infall
      and age, [Z/H], $\rm log(M_{\rm stellar}/M_{odot})$, $\log(\rm SFR$), TType
      and $\nabla$(g-i) of galaxy members of G and NG clusters. Galaxy
      members are split into bright (red/blue) or faint (green/pink). The curves and its width represent a cubic spline fitting to the median values and the variance calculated using the bootstrap technique, respectively.}
    \label{fig:tinfall_relation}
\end{figure*}

In this section, we explore in more detail the differences found in the
distribution of G and NG clusters on the PPS diagram, presented in
Section~\ref{sec:locus_properties}.
The delayed-then-rapid model \citep{2013MNRAS.432..336W}
proposes that galaxy quenching depends on $\rm t_{inf}$. We probe environmental effects on galaxies by studying the relation between $\rm t_{inf}$ and galaxy properties in G and NG
systems, separating galaxy populations into B and F. We consider that all galaxies in a single PNZ are well represented by the mean $\rm t_{inf}$ presented in P19. We show in Fig.~\ref{fig:tinfall_relation}, in panels (a) to (f), the median values of age, [Z/H], $\rm log(M_{stellar}/M_{\odot})$, log(SFR), TType and $\rm \nabla (g-i)$, respectively, for galaxies in each PNZ. In order to highlight global trends, we perform a spline fitting, which better shows the relation between infall time and galaxy properties. The curve width in Fig.~\ref{fig:tinfall_relation} represents the associated 1-$\sigma$ error in each PNZ. To address the statistical variance we use a bootstrap technique. We follow: 1) for each PNZ we randomly select N values (where N is the number of galaxies in the same region), with replacement, from the observed distribution; 2) calculate the variance using the new distribution $\rm Q_{sigma}$\footnote{The
  variance from quartiles is calculated as $\rm Q_{sigma} = 0.74
  \times (Q_{75\%} - Q_{25\%})$.}; 3) we repeat this procedure 1,000
times; and 4) consider the variance as the median of the
$\rm Q_{sigma}$ distribution. Additionally, we perform a similar analysis to the mass matched sample in the bright regime, in order to investigate how the mass mismatch between G and NG clusters (see Section \ref{sec:Structure}) affects our results. Quantitatively, we find mean differences\footnote{Over all PNZs for G and NG clusters.} of 0.06 Gyr, 0.001 dex, 0.01 dex, 0.02 dex, 0.05 and 0.03 for Age, [Z/H], log($\rm M_{stellar}/M_{\odot}$), log(SFR), TType and $\rm \nabla \, (g-i)$, respectively. Comparison with the uncertainties shown in Table \ref{table:starlight_errors} guarantees that the results shown in Fig. \ref{fig:infall_rate_estimate} do not depend on the use of mass-matched samples.

In a first approximation, we adopt a linear relation between $\rm
t_{inf}$ and galaxy properties. However, we emphasize that this is a
functional approach and the relations are not expected to be 
linear. Our goal here is to derive quantities such as
variations with infall time. We fit the observed relations via the
{\sc SciDAVis} statistical analysis tool \citep{benkert2014scidavis}, that takes into account errors in both x and y for the fit. The resulting
intercept and gradient coefficients are shown in Table~\ref{table:linear_coeficients}.
Column (1) lists cluster class
and luminosity regime; in column (2) we list the correspondent
coefficient; columns (3) to (8) show the results and associated errors
for age, [Z/H], log($\rm M_{stellar}/M_{\odot}$), log(SFR), TType and
$\rm \nabla(g-i)$, respectively. We compare the significance level of
the differences between G and NG clusters as:
\begin{equation}
  \rm  \sigma = \frac{\sqrt{(\sigma^{\chi}_{i,G})^{2} + (\sigma^{\chi}_{i,NG})^{2}}}{2},
\end{equation}
where $\rm \sigma^{\chi}_{i,G}$ and $\rm \sigma^{\chi}_{i,NG}$ are the errors
associated with the i-th coefficient (linear or angular) of the parameter
$\chi$ (age, for example). In Table~\ref{table:linear_coeficients}, we
highlight in red relevant differences in the gradients of G
and NG clusters. Dark red means differences greater than 2-$\sigma$ and
light red differences greater than 1-$\sigma$. We highlight similarly
differences in the intercept coefficients in blue color.

With respect to global trends, we see from Fig.~\ref{fig:tinfall_relation}
that the median stellar population parameters (panels a, b and c)
differ significantly between B and F galaxies, both in G and NG
systems. For instance, we find a mean difference of $2.2\pm 0.5$\,Gyr, $0.3 \pm 0.1$ and $0.99 \pm 0.06$\,dex between the intercept parameter of B and F galaxies for age,
[Z/H] and $\rm \log(M_{stellar}/M_{\odot})$, respectively. Additionally, the
relation with respect to SFR of B galaxies has a gradient closer to zero
than the one for F galaxies. We find an average
slope\footnote{Between G and NG  clusters.} of -0.08 and -0.20 for the SFR of B and F galaxies, respectively. These results unequivocally suggest that B and F
galaxies are distinctly affected by environmental quenching. However,
we find that indistinctly in G and NG clusters, both B and F galaxies
feature similar trends in stellar mass (within 0.1\,dex, comparable with
the measurement uncertainty), and the trend is
approximately constant with infall time. This result
may suggest that galaxy quenching is mostly related to the removal of the 
gas component instead of the stellar mass itself.

\begin{table*}
\centering
\caption{Gradient and Intercept Coefficients from the linear fit presented in Section~6.}
\label{table:linear_coeficients}

\resizebox{\textwidth}{!}{\begin{minipage}{\textwidth}
\centering

\begin{tabular}{c|c|cccccc}
\hline
                             & y = ax + b & Age (Gyr)                                 & [Z/H]                                      & $\rm log(M_{stellar}/M_{\odot})$         & log(SFR) ($\rm M_{\odot}/yr$)                & TType                                      & $\rm \nabla (g-i)$                          \\ \hline
                             & a          & \cellcolor[HTML]{FFCCC9}$0.701 \pm 0.052$ & $0.014 \pm 0.002$                          & {\color[HTML]{333333} $0.025 \pm 0.006$} & \cellcolor[HTML]{FFCCC9}$-0.078 \pm 0.011$  & $-0.333 \pm 0.042$                         & \cellcolor[HTML]{FFCCC9}$0.012 \pm 0.001$  \\
\multirow{-2}{*}{$\rm G_B$}  & b          & $4.468 \pm 0.219$                         & $-0.031 \pm  0.008$                        & $10.896 \pm  0.023$                      & \cellcolor[HTML]{DAE8FC}$-0.832 \pm 0.051$  & \cellcolor[HTML]{DAE8FC}$-0.266 \pm 0.201$ & \cellcolor[HTML]{34CDF9}$-0.115 \pm 0.006$ \\ \hdashline
                             & a          & \cellcolor[HTML]{FFCCC9}$0.538 \pm 0.105$ & $0.013 \pm 0.004$                          & $0.019 \pm 0.010$                        & \cellcolor[HTML]{FFCCC9}$-0.100 \pm 0.020$  & $-0.455 \pm 0.168$                         & \cellcolor[HTML]{FFCCC9}$0.003 \pm 0.006$  \\
\multirow{-2}{*}{$\rm NG_B$} & b          & $4.826 \pm 0.448$                         & $-0.022 \pm 0.014$                         & $10.925 \pm 0.041$                       & \cellcolor[HTML]{DAE8FC}$-0.706 \pm 0.089$  & \cellcolor[HTML]{DAE8FC}$0.553 \pm 0.643$  & \cellcolor[HTML]{34CDF9}$-0.078 \pm 0.009$ \\ \hline
                             & a          & \cellcolor[HTML]{FFCCC9}$0.626 \pm 0.068$ & \cellcolor[HTML]{FD6864}$0.052 \pm 0.010$  & $0.037 \pm 0.012$                        & \cellcolor[HTML]{FD6864}$-0.249 \pm 0.027 $ & $-0.611 \pm 0.151$                         & $0.009 \pm 0.004$                          \\
\multirow{-2}{*}{$\rm G_F$}  & b          & \cellcolor[HTML]{DAE8FC}$2.043 \pm 0.705$ & \cellcolor[HTML]{34CDF9}$-0.360 \pm 0.040$ & $9.973 \pm 0.043$                        & \cellcolor[HTML]{34CDF9}$-0.139 \pm 0.110$  & $2.497 \pm 0.661$                          & \cellcolor[HTML]{DAE8FC}$-0.080 \pm 0.020$ \\ \hdashline
                             & a          & \cellcolor[HTML]{FFCCC9}$0.511 \pm 0.101$ & \cellcolor[HTML]{FD6864}$0.021 \pm 0.010$  & $0.035 \pm 0.012$                        & \cellcolor[HTML]{FD6864}$-0.141 \pm 0.038$  & $-0.755 \pm 0.079$                         & $0.005 \pm 0.002$                          \\
\multirow{-2}{*}{$\rm NG_F$} & b          & \cellcolor[HTML]{DAE8FC}$2.888 \pm 0.446$ & \cellcolor[HTML]{34CDF9}$-0.192 \pm 0.039$ & $9.928 \pm 0.046$                        & \cellcolor[HTML]{34CDF9}$-0.692 \pm 0.162$  & $3.257 \pm 0.275$                          & \cellcolor[HTML]{DAE8FC}$-0.057 \pm 0.010$ \\ \hline
\end{tabular}
\end{minipage}}
\end{table*}

In panel (a) of Fig.~\ref{fig:tinfall_relation}, two noticeable
trends can be found regarding the evolution of galaxy properties
with $\rm t_{inf}$ in G and NG clusters. First, $\rm G_B$ galaxies with
$\rm t_{inf} \geq 3.80$\,Gyr are on average $0.3\pm 0.2$ \,Gyr older than $\rm NG_B$
systems. Furthermore, we find that at 
$\rm t_{inf} < 4.8$\,Gyr $\rm NG_F$ galaxies are on average $0.6 \pm 0.5$\,Gyr
older than those in $\rm G_F$. The third column of Table~\ref{table:linear_coeficients} 
shows that in both bright and faint regimes, the slope of the
relation between age and $\rm t_{inf}$ in NG clusters lies more than
1-$\sigma$ lower than that for G clusters. In panel (b) we see that [Z/H]
behaves similarly. Despite no significant difference for B
galaxies, F galaxies with $\rm t_{inf} < 4.8$\,Gyr in NG clusters are $0.05\pm0.02$\,dex 
more metal-rich than in G clusters. We also note a difference greater
than 2-$\sigma$ in the slope of the [Z/H] relation in G and NG
clusters. Namely, we find a shallower relation in NG clusters compared
to G systems (see Table~\ref{table:linear_coeficients}, fourth
column). In other words, the trends found in panels (a) and (b)
suggest that F galaxies with $\rm 4.5 < t_{inf} < 1.5$ Gyr in NG clusters are older and more metal-rich than in G clusters. On the other hand, the excess of B galaxies that have younger ages at high infall time in NG clusters may
suggest that G clusters have a better defined virialized core, as we
will discuss in Section~\ref{sec:Discussion}. In panel (c), we do see
that $\rm M_{stellar}$ is approximately constant with infall time,
indistinctly of cluster class and luminosity regime. This translates
to slopes lower than 0.1 dex in all cases. In panel (d), the 
SFR is shown as a function of infall time. SFR behaves similarly to [Z/H].
In the bright regime, we note that galaxies do not show significant
differences in SFR in G and NG clusters. However, the SFR
slope for G and NG clusters differs by more than 1-$\sigma$. These 
differences are more evident in the faint regime: at $\rm t_{inf} < 4.8$\,Gyr 
$\rm NG_F$ galaxies are less star-forming by, on average, $0.21 \pm 0.17$\,dex
than $\rm G_F$ systems. The slope of the SFR relation differs
more than 2-$\sigma$ between G and NG clusters. Similarly to age and 
[Z/H], NG clusters feature shallower slopes in comparison to G clusters, therefore implying
a weak dependence with infall time. Panel~(e) shows the trends with
morphology (TType), appearing very similar to those for the SFR,
supporting the relation between star
formation quenching and morphological transition. F galaxies with
$\rm t_{inf} < 4.8$\,Gyr in NG clusters have lower values of TType, $0.27 \pm 0.09$
on average, in comparison to G clusters. Finally, panel (f) plots 
the relation between $\rm \nabla (g-i)$ and infall time. 
Significant differences are found in both B and F subsets. In the bright
regime, $\rm G_B$ galaxies show a constantly
increasing (negative) color gradient with infall time, while $\rm NG_B$ galaxies
show a plateau after $\rm t_{inf} \sim 2.5$\,Gyr. The slopes of
these two relations differ more than 1-$\sigma$, as can be seen in
Table~\ref{table:linear_coeficients}. In the faint regime, we note that
$\rm NG_F$ galaxies have unequivocally shallower color gradients
than $\rm G_F$ systems. Quantitatively, we find a mean difference of 0.026,
being more negative in NG clusters.




\section{An Estimate of the Infalling Rate in NG Clusters}
\label{sec:infall_rate_estimate}

Several works relate the Non-Gaussianity of the velocity distribution
with a higher infall rate in NG clusters compared to G systems (e.g. \citealt{2017MNRAS.467.3268R}, \citealt{2017AJ....154...96D}). In
this section, we present further evidence of a higher infall rate in NG
systems by detailing the distribution of stellar mass in the
PPS. Galaxies in the PNZ 8 (see Table 1 in P19) have an average infall time of 1.42\,Gyr and
are mainly first infallers. However, the discretization of the PPS
presented by P19 is limited to $\rm R_{200}$ and it is expected that
galaxies first infalling in clusters may be also found beyond this
threshold. Thus we decided to include the Rhee Region~A in order to
account for galaxies beyond $\rm R_{200}$ (see Fig.~6 of R17). This region is mostly occupied
by interlopers. However, it is expected that the Shiftgapper technique
returns a catalog of bona fide members. Hence, we consider that
galaxies in the Rhee Region A correspond to the second most probable
population, namely first infallers.

Differently from the previous analysis, here we consider the PPS for
each cluster separately. We calculate the sum of stellar mass in PNZ 8 (or Rhee Region A) for each cluster and then take an average
value for a given cluster class (G or NG) and luminosity regime (B or
F). In the bright regime, 
NG clusters have an excess of stellar mass of 
$\rm 0.51\times 10^{11}\,M_{\odot}$ (PNZ 8) and
$\rm 0.84 \times 10^{12}\,M_{\odot}$ (Rhee A), 
with respect to G clusters. In the faint regime, we also
note NG clusters have an excess of $\rm 0.33 \times 10^{11}M_{\odot}$ (PNZ 8) and
$\rm 0.31 \times 10^{12} M_{\odot}$ (Rhee A) with respect to 
G clusters. Putting together the contributions of B
and F galaxies, we find that NG clusters have an excess of
$\rm \sim 10^{11} M_{\odot}$ in the PNZ 8 and $\rm \sim 10^{12} M_{\odot}$ in the
Rhee Region A. This unambiguously shows that there are more galaxies
infalling in NG clusters in comparison to G clusters.

\begin{figure}
    \centering
    \includegraphics[width = \columnwidth]{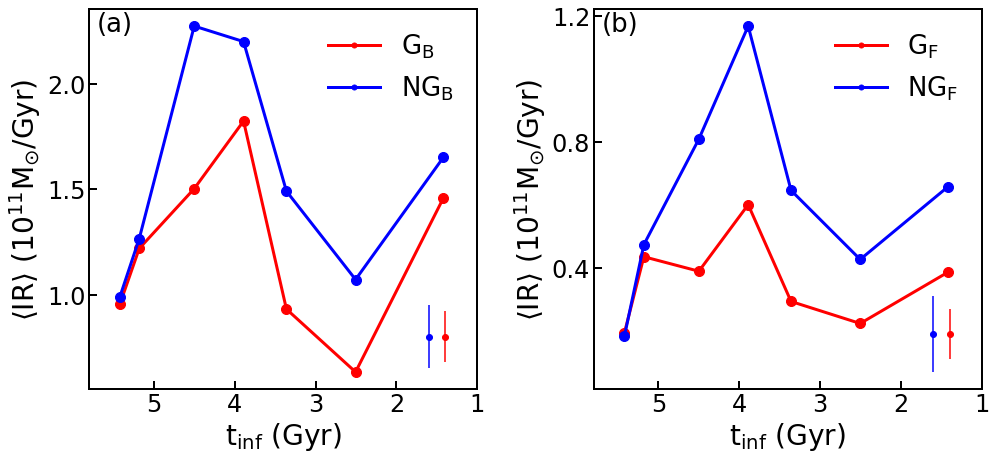}
    \caption{Estimate of mean infall rate of B (left) and F (right)
      luminosity regime as a function of $t_{\rm inf}$ for G and NG
      clusters. Non-Gaussian systems show higher values of infall rate
      in both, with respect to  G systems.}
    \label{fig:infall_rate_estimate}
\end{figure}

Using the relation between locus in the PPS and infall time we can
derive a rough estimate of the infall rate in NG clusters. We calculate 
the mean infall rate ($\rm \langle IR \rangle$) as follows: 1) For
each cluster we sum the stellar mass within a single PNZ and divide it by
the mean infall range of PNZ i - PNZ i-1; 2) we take the average
value for each PNZ for a given cluster class and luminosity regime and
3) these estimates across all PNZ regions provide an infall history for G and NG
clusters. They represent a rough estimate of the amount of stellar mass accreted into clusters from the infall time of PNZ i-1 to the PNZ i, with i varying from 1 to 8\footnote{At i = 8, the infall range is from 0 to $\rm t_{inf}(PNZ 8)$}  The results are shown in Fig.~\ref{fig:infall_rate_estimate}, where panels (a) and (b) correspond to B and F galaxies, respectively. We show at the bottom right of each panel the mean error for each case. We note that NG clusters feature 
larger stellar mass across all values of infall time. We find a mean
difference of $\rm \langle \langle IR \rangle_{NG} - \langle IR \rangle_{G} \rangle = 0.4 \times 10^{11}M_{\odot}$ and $\rm 0.3 \times 10^{11}M_{\odot}$ in the B and F regimes, respectively. An integration of the relation between $\rm \langle IR \rangle$ and $\rm t_{inf}$ suggests that NG clusters accreted $\rm (1.5 \pm 0.8) \times 10^{12} M_{\odot}$ more stellar mass in the last $\sim 5\,Gyr$ than G systems. For comparison, it roughly corresponds to the stellar mass of the local group, $\rm \sim 10^{12} M_{\odot}$.

A potential sample bias caveat relates to the different virial mass
distributions of G and NG clusters, as shown in panel (a) of
Fig.~\ref{fig:concentration}.  In order to guarantee that NG clusters
have a higher infall rate regardless of their mass, we separate our
sample in fixed bins in log($\rm M_{200}/M_{\odot}$) from 14 to 14.75 in
steps of 0.25. The bin size is roughly half of the error ($\sim 0.13$ dex) in $\rm log(M_{200}/M_{\odot})$ \citep[dC17]{2007ApJ...671..153Y}. This guarantees that the mass distribution within each bin has no significant differences between G and NG clusters. The [14,14.75] range is chosen due to a limitation of the
faint component of G clusters. Namely, we do not find G clusters
in the F regime with log($\rm M_{200}/M_{\odot} > 14.75$).
Thus this range guarantees that
we have G and NG clusters in every bin for both luminosity
regimes. The analysis reveals that, in all three bins, NG clusters have
accreted more mass over the last $\sim 5$\,Gyr in comparison to G
clusters. Taking the average over the three virial mass bins we find
that NG clusters accreted an excess of $ \rm (1.4 \pm 0.7) \times 10^{11}
M_{\odot}$ and $\rm (0.9 \pm 0.5) \times 10^{11} M_{\odot}$ in the B and
F regimes, respectively. These trends show that most of the difference
originates in the tail end of the halo distribution, but 
confirms that the NG classification is directly related to a higher
infall rate regardless of virial mass.

\section{Discussion and Conclusions}
\label{sec:Discussion}

\subsection{G and NG Clusters: Different Environments to Study Galaxy Evolution}

A comparison of G and NG clusters indicates that these two groups have
significant differences in both structure and galaxy properties.
We summarize those differences below:

\begin{itemize}
\item NG Clusters are more massive, larger (i.e. greater $\rm R_{200})$, less concentrated and show an excess of fainter galaxies with respect to G clusters (Fig.~\ref{fig:concentration});

\item Fig.~\ref{fig:PPS_Density} highlights that higher density regions in the PPS
  of NG clusters extend to higher velocities in comparison to G Clusters, a possible
  signature from an excess of high-velocity infalling galaxies;

\item The distribution of galaxy properties is systematically more mixed in
  the PPS of NG clusters than in G clusters (Figs.~\ref{fig:full_grid} and \ref{fig:grid_mstellar}).
  This trend holds in both luminosity regimes, but it is more noticeable in F galaxies;
    
\item Fig.~\ref{fig:infall_rate_estimate} suggests that G and NG clusters have
  different accretion histories over the last $\sim 5$\,Gyr. Our findings suggest
  that NG clusters accreted on average $\rm \sim 10^{11} M_{\odot}$ more stellar mass than G systems. The trend of a higher accretion rate in NG clusters is also present when
  comparing systems with similar $\rm M_{200}$;
\end{itemize}

These results suggest that G and NG clusters provide two different
environments to study galaxy evolution and environmental
quenching. N-body simulations show that more massive clusters sustain
higher merger rates \citep{2009ApJ...701.2002G}. In other words,
massive clusters are usually formed by the interaction of two groups
and/or clusters. Our sample is built upon massive clusters and hence
it is expected that a large fraction of them had experienced such
fusion events in the past. A higher fraction of merging events is
expected in NG clusters due to the observed excess of massive systems
in comparison to G groups. Such events are perturbations in the
dynamical equilibrium of a cluster. From the PPS point of view, this
translates into a more mixed distribution of galaxy properties in
comparison to the unperturbed state. Our results point to NG clusters
having more mixed distributions in the PPS. Finally,
\cite{2019MNRAS.490..773R} show that NG clusters suffered their last
major merger more recently than G clusters. These trends also show that NG
clusters are statistically in a more unrelaxed state in comparison to
G clusters and thus provide different environments to study galaxy
evolution.

Previous works indicate that NG clusters also show higher infall
rates. Here we quantify this infall rate by tracing the stellar mass in
different regions of the PPS and estimate that over the last 5\,Gyr of
cosmic time, NG clusters roughly accreted stellar material comparable to the stellar component of massive groups, in comparison to G systems. At first, approximately 2/3 of the accreted mass comes from B galaxies. However, B galaxies are roughly 10 times more massive than F galaxies. This means that numerically there are far more F galaxies being accreted in comparison to B galaxies. We conclude that this is mostly due to the accretion of low-luminosity galaxies (FG), a result consistent with works that explain the non-gaussianity in the projected velocity distribution as a result of an accretion of low luminosity galaxies
\citep[e.g.][]{2017AJ....154...96D,2019MNRAS.490..773R}. Examining
Table~\ref{table:linear_coeficients}, we note that NG clusters feature
shallower color gradients (i.e. slopes closer to zero) in comparison to G
clusters. These gradients reflect a more mixed galactic population in
NG systems. However, we highlight that these results are found by
considering large samples of clusters to guarantee that the effects of
projection along the line of sight influence both cluster classes in
the same way. \cite{2018MNRAS.473L..31C} study the galaxy velocity
dispersion and conclude that B and F galaxies are more distinct in G
clusters. It follows from the mass-segregation phenomenon in clusters
\citep{1980ApJ...241..521C} that relaxed systems show a clear
separation between high and low mass galaxy properties, hence a more
mixed population of bright and faint galaxies gives further support
to the hypothesis that NG clusters are in a more unrelaxed state than G
systems. Here we also provide further evidence that the
non-gaussianity in the projected velocity distribution is connected to
a higher infall rate of faint galaxies, in agreement with
\cite{2019MNRAS.490..773R}. However, a more detailed characterization
on a one-by-one basis should be given by combining
further dynamic probes such as caustic curve analysis
\citep{gifford2013systematic,2012ApJ...747L..42D}, X-Ray
characterization \citep{2001A&A...378..408S}, velocity dispersion
profiles \citep{2019MNRAS.482.5138B} and galaxy spatial distribution
\citep{2006A&A...450....9F}.

\subsection{Low \emph{vs.} High Mass Galaxies Quenching in Dense Environments}

One of the primary results of this work is the unequivocal trends
found between G and NG clusters and the properties of the galaxy
members, shown in Fig.~\ref{fig:tinfall_relation}. In this regard,
we highlight the following points:

\begin{itemize}
\item The trends of age, [Z/H], SFR and TType of B galaxies with
  respect to infall time show slopes closer to zero in comparison to F
  galaxies, indistinctly in G or NG clusters. These trends evidence
  how B and F galaxies are differently affected by their
  environment. Namely, B galaxies are more massive and hence less
  affected by environment-specific processes such as ram pressure stripping;
        
\item Age and [Z/H] have very similar relations with infall time,
  reinforcing the closeness between these two parameters. (Luminosity-weighted) Age
  here roughly means the time since the last star formation episode and thus
  higher age translates to more time to stars evolve and increase the
  metallicity of the ISM;
        
\item The similarity between the relations of SFR and TType
  suggests that quenching of star formation and morphological
  transformations may be simultaneously caused by a third process;
        
\item The slopes of the trends with infall time regarding FG are
  consistently lower in NG clusters in comparison to G systems, for
  most of the six-parameter set here adopted to characterize member
  galaxy properties;
    
\item F galaxies with time since infall lower than 4.5\,Gyr in G
  clusters are younger, more metal-poor and show higher star formation rates 
  than their counterparts in NG clusters;
\end{itemize}

These relations are of particular interest since the quenching time
scale is pivotal to understanding environmental effects in galaxy
evolution \citep{2016ApJ...816L..25P}. Galaxies moving within and/or
towards the cluster eventually reach a threshold density that triggers
efficient ram pressure stripping, removing the gas component of low
mass galaxies in a short time scale ($\leq$ 1\,Gyr, \citealt{2007MNRAS.380.1399R}) quickly quenching
star formation (R19). However, the trends for high mass
galaxies seems to be quite different. The flatter relations between
[Z/H], SFR or TType and $\rm t_{inf}$ for B galaxies in comparison to
F ones indicate that massive (and hence more luminous) galaxies are
less affected by the environment, indistinctly in G or NG
clusters. This result is in agreement with massive galaxies being
quenched mostly due to internal mechanisms such as AGN and stellar
feedback \citep{1974MNRAS.169..229L}. Various works debate the relation between galaxy star formation quenching and morphological transformation \citep{2009ApJ...707..250M,2014MNRAS.440..889S,2019MNRAS.486..868K}. The trends found in Fig.~\ref{fig:tinfall_relation} support the close SFR - morphology relation and extends it also to age and [Z/H], providing further
insight into how stellar population parameters reflect galaxy evolution. 
 
Additionally, in the faint regime, we also note striking
differences between G and NG clusters, suggesting once more that bright
galaxies are likely quenched upon internal processes. We find that galaxies with $\rm 2.5<t_{inf} < 4.5$\,Gyr in NG systems are older and more metal-rich than their counterparts in G systems. This corresponds roughly to PNZs 4 to 6, which are characterized by a mean $\rm |V_{LOS}|/\sigma \sim 1.2 \pm 0.3 \, km\,s^{-1}$. A higher $\rm |V_{LOS}|/\sigma$ is closely connected to infalling galaxies. Namely, galaxies with higher velocities are those first entering the cluster. The trends we find hence indicate that galaxies infalling in NG clusters are older and more metal-rich than in G systems. These results are in agreement with dC17, which find similar trends when comparing radial profiles of galaxy properties in G and NG clusters. Furthermore, despite our analysis being limited to $\rm R_{200}$, dC17 show that these differences extend until 2$\rm R_{200}$. These results are in agreement with a scenario where galaxies infalling into NG clusters have been pre-processed (P19) in comparison to those in G clusters. This is reinforced by the results found for SFR, TType and color gradient,
namely we find galaxies with lower SFR and TType in NG systems. Also,
faint galaxies in NG clusters show shallower color gradients
than faint galaxies in G systems, indicating that
galaxies entering NG clusters already started a 
transformation towards an elliptical morphology. Complementary to the results
presented in this paper, \cite{2019MNRAS.487L..86D} show that faint spiral
galaxies are the ones that suffer major environmental effects.

This paper is the fourth of a series where we investigate how the gaussianity of the velocity distribution of the galaxy members in a cluster is of paramount importance for the studying of how galaxy evolution is affected by the environment. NG clusters exhibit a higher infall rate when contrasted to the G ones. Previous works have suggested this trend, here we quantify it for the first time. Also, based on the PPS analysis, we show that galaxies with high $\rm |V_{LOS}|/\sigma$ belonging to NG clusters are older and more metal-rich than the ones in G systems (even for $\rm R < R_{200}$). This is a clear indication that these galaxies infalling into NG systems were preprocessed from the chemical evolution standpoint. Examining the trends of Age, [Z/H], SFR, $\rm M_{stellar}$, TType and $\nabla$(g-i), with the infall time, for faint galaxies in NG versus G, confirms that such a distinction in velocity distribution is not an artifact of the methods employed to measure it.

\section*{Acknowledgements}

We would like to thank the anonymous referee for the comments and
suggestions which contributed to greatly improving this article. RRdC and VMS thank T. S. Golçalves for fruitful discussions on this topic. VMS
acknowledges the CAPES scholarship through the grants
88887.508643/2020-00 and 88882.444468/2019-01. RRdC aknowledges the
financial support from FAPESP through the grant \#2014/11156-4.  TFL
acknowledges financial support from FAPESP (2018/02626-8) and CNPq
(306163/2019-5). IF acknowledges support from the Spanish Ministry of
Science, Innovation and Universities (MCIU), through grant
PID2019-104788GB-I00. ALBR thanks the support of CNPq, grant 311932/2017-7. S.B.R. acknowledges support from Conselho Nacional de Desenvolvimento Científico e Tecnológico – CNPq. This work was made possible thanks to a number of
open-source software packages: AstroPy \citep{2018AJ....156..123A},
Matplotib \citep{2005ASPC..347...91B}, NumPy
\citep{2011CSE....13b..22V}, Pandas \citep{mckinney2010data} and SciPy
\citep{2020zndo....595738V}.

\section*{Data Availability}
The data used in this manuscript will be made available under request.




\bibliographystyle{mnras}
\bibliography{bib} 

\begin{thebibliography}{}
\makeatletter
\relax
\def\mn@urlcharsother{\let\do\@makeother \do\$\do\&\do\#\do\^\do\_\do\%\do\~}
\def\mn@doi{\begingroup\mn@urlcharsother \@ifnextchar [ {\mn@doi@}
  {\mn@doi@[]}}
\def\mn@doi@[#1]#2{\def\@tempa{#1}\ifx\@tempa\@empty \href
  {http://dx.doi.org/#2} {doi:#2}\else \href {http://dx.doi.org/#2} {#1}\fi
  \endgroup}
\def\mn@eprint#1#2{\mn@eprint@#1:#2::\@nil}
\def\mn@eprint@arXiv#1{\href {http://arxiv.org/abs/#1} {{\tt arXiv:#1}}}
\def\mn@eprint@dblp#1{\href {http://dblp.uni-trier.de/rec/bibtex/#1.xml}
  {dblp:#1}}
\def\mn@eprint@#1:#2:#3:#4\@nil{\def\@tempa {#1}\def\@tempb {#2}\def\@tempc
  {#3}\ifx \@tempc \@empty \let \@tempc \@tempb \let \@tempb \@tempa \fi \ifx
  \@tempb \@empty \def\@tempb {arXiv}\fi \@ifundefined
  {mn@eprint@\@tempb}{\@tempb:\@tempc}{\expandafter \expandafter \csname
  mn@eprint@\@tempb\endcsname \expandafter{\@tempc}}}

\bibitem[\protect\citeauthoryear{{Abadi}, {Moore}  \& {Bower}}{{Abadi}
  et~al.}{1999}]{1999MNRAS.308..947A}
{Abadi} M.~G.,  {Moore} B.,   {Bower} R.~G.,  1999, \mn@doi [\mnras]
  {10.1046/j.1365-8711.1999.02715.x}, \href
  {https://ui.adsabs.harvard.edu/abs/1999MNRAS.308..947A} {308, 947}

\bibitem[\protect\citeauthoryear{{Adelman-McCarthy} et~al.,}{{Adelman-McCarthy}
  et~al.}{2007}]{2007ApJS..172..634A}
{Adelman-McCarthy} J.~K.,  et~al., 2007, \mn@doi [\apjs] {10.1086/518864},
  \href {https://ui.adsabs.harvard.edu/abs/2007ApJS..172..634A} {172, 634}

\bibitem[\protect\citeauthoryear{{Albareti} et~al.,}{{Albareti}
  et~al.}{2017}]{2017ApJS..233...25A}
{Albareti} F.~D.,  et~al., 2017, \mn@doi [\apjs] {10.3847/1538-4365/aa8992},
  \href {https://ui.adsabs.harvard.edu/abs/2017ApJS..233...25A} {233, 25}

\bibitem[\protect\citeauthoryear{{Anderson} \& {Darling}}{{Anderson} \&
  {Darling}}{1952}]{anderson1952asymptotic}
{Anderson} T.~W.,  {Darling} D.~A.,  1952, The annals of mathematical
  statistics, pp 193--212

\bibitem[\protect\citeauthoryear{{Angthopo}, {Ferreras}  \& {Silk}}{{Angthopo}
  et~al.}{2019}]{2019MNRAS.488L..99A}
{Angthopo} J.,  {Ferreras} I.,   {Silk} J.,  2019, \mn@doi [\mnras]
  {10.1093/mnrasl/slz106}, \href
  {https://ui.adsabs.harvard.edu/abs/2019MNRAS.488L..99A} {488, L99}

\bibitem[\protect\citeauthoryear{{Astropy Collaboration} et~al.,}{{Astropy
  Collaboration} et~al.}{2018}]{2018AJ....156..123A}
{Astropy Collaboration} et~al., 2018, \mn@doi [\aj] {10.3847/1538-3881/aabc4f},
  \href {https://ui.adsabs.harvard.edu/abs/2018AJ....156..123A} {156, 123}

\bibitem[\protect\citeauthoryear{{Balogh}, {Navarro}  \& {Morris}}{{Balogh}
  et~al.}{2000}]{2000ApJ...540..113B}
{Balogh} M.~L.,  {Navarro} J.~F.,   {Morris} S.~L.,  2000, \mn@doi [\apj]
  {10.1086/309323}, \href
  {https://ui.adsabs.harvard.edu/abs/2000ApJ...540..113B} {540, 113}

\bibitem[\protect\citeauthoryear{{Barrett}, {Hunter}, {Miller}, {Hsu}  \&
  {Greenfield}}{{Barrett} et~al.}{2005}]{2005ASPC..347...91B}
{Barrett} P.,  {Hunter} J.,  {Miller} J.~T.,  {Hsu} J.~C.,   {Greenfield} P.,
  2005, in {Shopbell} P.,  {Britton} M.,   {Ebert} R.,  eds,  Astronomical
  Society of the Pacific Conference Series Vol. 347, Astronomical Data Analysis
  Software and Systems XIV. p.~91

\bibitem[\protect\citeauthoryear{{Baumgardt}, {Hilker}, {Sollima}  \&
  {Bellini}}{{Baumgardt} et~al.}{2019}]{2019MNRAS.482.5138B}
{Baumgardt} H.,  {Hilker} M.,  {Sollima} A.,   {Bellini} A.,  2019, \mn@doi
  [\mnras] {10.1093/mnras/sty2997}, \href
  {https://ui.adsabs.harvard.edu/abs/2019MNRAS.482.5138B} {482, 5138}

\bibitem[\protect\citeauthoryear{Benkert, Franke, Pozitron  \&
  Standish}{Benkert et~al.}{2014}]{benkert2014scidavis}
Benkert T.,  Franke K.,  Pozitron D.,   Standish R.,  2014, D005 (Free Software
  Foundation, Inc: 51 Franklin Street, Fifth Floor, Boston, MA 02110-1301 USA)

\bibitem[\protect\citeauthoryear{{Blanton} et~al.,}{{Blanton}
  et~al.}{2005}]{2005AJ....129.2562B}
{Blanton} M.~R.,  et~al., 2005, \mn@doi [\aj] {10.1086/429803}, \href
  {https://ui.adsabs.harvard.edu/abs/2005AJ....129.2562B} {129, 2562}

\bibitem[\protect\citeauthoryear{{Bongiorno} et~al.,}{{Bongiorno}
  et~al.}{2016}]{Bongiorno}
{Bongiorno} A.,  et~al., 2016, \mn@doi [\aap] {10.1051/0004-6361/201527436},
  \href {https://ui.adsabs.harvard.edu/abs/2016A&A...588A..78B} {588, A78}

\bibitem[\protect\citeauthoryear{{Brinchmann}, {Charlot}, {White}, {Tremonti},
  {Kauffmann}, {Heckman}  \& {Brinkmann}}{{Brinchmann}
  et~al.}{2004}]{2004MNRAS.351.1151B}
{Brinchmann} J.,  {Charlot} S.,  {White} S.~D.~M.,  {Tremonti} C.,  {Kauffmann}
  G.,  {Heckman} T.,   {Brinkmann} J.,  2004, \mn@doi [\mnras]
  {10.1111/j.1365-2966.2004.07881.x}, \href
  {https://ui.adsabs.harvard.edu/abs/2004MNRAS.351.1151B} {351, 1151}

\bibitem[\protect\citeauthoryear{{Capelato}, {Gerbal}, {Salvador-Sole},
  {Mathez}, {Mazure}  \& {Sol}}{{Capelato} et~al.}{1980}]{1980ApJ...241..521C}
{Capelato} H.~V.,  {Gerbal} D.,  {Salvador-Sole} E.,  {Mathez} G.,  {Mazure}
  A.,   {Sol} H.,  1980, \mn@doi [\apj] {10.1086/158366}, \href
  {https://ui.adsabs.harvard.edu/abs/1980ApJ...241..521C} {241, 521}

\bibitem[\protect\citeauthoryear{{Choi}, {Han}  \& {Kim}}{{Choi}
  et~al.}{2010}]{2010JKAS...43..191C}
{Choi} Y.-Y.,  {Han} D.-H.,   {Kim} S.~S.,  2010, \mn@doi [Journal of Korean
  Astronomical Society] {10.5303/JKAS.2010.43.6.191}, \href
  {https://ui.adsabs.harvard.edu/abs/2010JKAS...43..191C} {43, 191}

\bibitem[\protect\citeauthoryear{{Cid Fernandes}, {Mateus}, {Sodr{\'e}},
  {Stasi{\'n}ska}  \& {Gomes}}{{Cid Fernandes}
  et~al.}{2005}]{2005MNRAS.358..363C}
{Cid Fernandes} R.,  {Mateus} A.,  {Sodr{\'e}} L.,  {Stasi{\'n}ska} G.,
  {Gomes} J.~M.,  2005, \mn@doi [\mnras] {10.1111/j.1365-2966.2005.08752.x},
  \href {https://ui.adsabs.harvard.edu/abs/2005MNRAS.358..363C} {358, 363}

\bibitem[\protect\citeauthoryear{{Costa}, {Ribeiro}  \& {de Carvalho}}{{Costa}
  et~al.}{2018}]{2018MNRAS.473L..31C}
{Costa} A.~P.,  {Ribeiro} A.~L.~B.,   {de Carvalho} R.~R.,  2018, \mn@doi
  [\mnras] {10.1093/mnrasl/slx156}, \href
  {https://ui.adsabs.harvard.edu/abs/2018MNRAS.473L..31C} {473, L31}

\bibitem[\protect\citeauthoryear{{Cox}, {Jonsson}, {Somerville}, {Primack}  \&
  {Dekel}}{{Cox} et~al.}{2008}]{2008MNRAS.384..386C}
{Cox} T.~J.,  {Jonsson} P.,  {Somerville} R.~S.,  {Primack} J.~R.,   {Dekel}
  A.,  2008, \mn@doi [\mnras] {10.1111/j.1365-2966.2007.12730.x}, \href
  {https://ui.adsabs.harvard.edu/abs/2008MNRAS.384..386C} {384, 386}

\bibitem[\protect\citeauthoryear{{Dawson} et~al.,}{{Dawson}
  et~al.}{2012}]{2012ApJ...747L..42D}
{Dawson} W.~A.,  et~al., 2012, \mn@doi [\apjl] {10.1088/2041-8205/747/2/L42},
  \href {https://ui.adsabs.harvard.edu/abs/2012ApJ...747L..42D} {747, L42}

\bibitem[\protect\citeauthoryear{De~Helguero}{De~Helguero}{1904}]{HEL1904}
De~Helguero Roma D. D.~F.,  1904, \mn@doi [Biometrika] {10.1093/biomet/3.1.84},
  3, 84

\bibitem[\protect\citeauthoryear{{Dekel} \& {Silk}}{{Dekel} \&
  {Silk}}{1986}]{1986ApJ...303...39D}
{Dekel} A.,  {Silk} J.,  1986, \mn@doi [\apj] {10.1086/164050}, \href
  {https://ui.adsabs.harvard.edu/abs/1986ApJ...303...39D} {303, 39}

\bibitem[\protect\citeauthoryear{{Dom{\'\i}nguez S{\'a}nchez},
  {Huertas-Company}, {Bernardi}, {Tuccillo}  \& {Fischer}}{{Dom{\'\i}nguez
  S{\'a}nchez} et~al.}{2018}]{2018MNRAS.476.3661D}
{Dom{\'\i}nguez S{\'a}nchez} H.,  {Huertas-Company} M.,  {Bernardi} M.,
  {Tuccillo} D.,   {Fischer} J.~L.,  2018, \mn@doi [\mnras]
  {10.1093/mnras/sty338}, \href
  {https://ui.adsabs.harvard.edu/abs/2018MNRAS.476.3661D} {476, 3661}

\bibitem[\protect\citeauthoryear{{Dressler}}{{Dressler}}{1980}]{Dressler}
{Dressler} A.,  1980, \mn@doi [\apj] {10.1086/157753}, \href
  {https://ui.adsabs.harvard.edu/abs/1980ApJ...236..351D} {236, 351}

\bibitem[\protect\citeauthoryear{{Dressler} \& {Shectman}}{{Dressler} \&
  {Shectman}}{1988}]{1988AJ.....95..985D}
{Dressler} A.,  {Shectman} S.~A.,  1988, \mn@doi [\aj] {10.1086/114694}, \href
  {https://ui.adsabs.harvard.edu/abs/1988AJ.....95..985D} {95, 985}

\bibitem[\protect\citeauthoryear{Engmann \& Cousineau}{Engmann \&
  Cousineau}{2011}]{engmann2011comparing}
Engmann S.,  Cousineau D.,  2011, Journal of applied quantitative methods, 6, 1

\bibitem[\protect\citeauthoryear{{Fadda}, {Girardi}, {Giuricin}, {Mardirossian}
   \& {Mezzetti}}{{Fadda} et~al.}{1996}]{1996ApJ...473..670F}
{Fadda} D.,  {Girardi} M.,  {Giuricin} G.,  {Mardirossian} F.,   {Mezzetti} M.,
   1996, \mn@doi [\apj] {10.1086/178180}, \href
  {https://ui.adsabs.harvard.edu/abs/1996ApJ...473..670F} {473, 670}

\bibitem[\protect\citeauthoryear{{Feigelson} \& {Babu}}{{Feigelson} \&
  {Babu}}{2012}]{2012msma.book.....F}
{Feigelson} E.~D.,  {Babu} G.~J.,  2012, {Modern Statistical Methods for
  Astronomy}

\bibitem[\protect\citeauthoryear{{Flin} \& {Krywult}}{{Flin} \&
  {Krywult}}{2006}]{2006A&A...450....9F}
{Flin} P.,  {Krywult} J.,  2006, \mn@doi [\aap] {10.1051/0004-6361:20041635},
  \href {https://ui.adsabs.harvard.edu/abs/2006A&A...450....9F} {450, 9}

\bibitem[\protect\citeauthoryear{{Fujita}}{{Fujita}}{2004}]{2004PASJ...56...29F}
{Fujita} Y.,  2004, \mn@doi [\pasj] {10.1093/pasj/56.1.29}, \href
  {https://ui.adsabs.harvard.edu/abs/2004PASJ...56...29F} {56, 29}

\bibitem[\protect\citeauthoryear{Gehan}{Gehan}{1965}]{gehan1965generalized}
Gehan E.~A.,  1965, Biometrika, 52, 203

\bibitem[\protect\citeauthoryear{{Genel}, {Genzel}, {Bouch{\'e}}, {Naab}  \&
  {Sternberg}}{{Genel} et~al.}{2009}]{2009ApJ...701.2002G}
{Genel} S.,  {Genzel} R.,  {Bouch{\'e}} N.,  {Naab} T.,   {Sternberg} A.,
  2009, \mn@doi [\apj] {10.1088/0004-637X/701/2/2002}, \href
  {https://ui.adsabs.harvard.edu/abs/2009ApJ...701.2002G} {701, 2002}

\bibitem[\protect\citeauthoryear{Gifford, Miller  \& Kern}{Gifford
  et~al.}{2013}]{gifford2013systematic}
Gifford D.,  Miller C.,   Kern N.,  2013, The Astrophysical Journal, 773, 116

\bibitem[\protect\citeauthoryear{{Girardi}, {Escalera}, {Fadda}, {Giuricin},
  {Mardirossian}  \& {Mezzetti}}{{Girardi} et~al.}{1997}]{1997ApJ...482...41G}
{Girardi} M.,  {Escalera} E.,  {Fadda} D.,  {Giuricin} G.,  {Mardirossian} F.,
   {Mezzetti} M.,  1997, \mn@doi [\apj] {10.1086/304113}, \href
  {https://ui.adsabs.harvard.edu/abs/1997ApJ...482...41G} {482, 41}

\bibitem[\protect\citeauthoryear{{Gunn} \& {Gott}}{{Gunn} \&
  {Gott}}{1972}]{1972ApJ...176....1G}
{Gunn} J.~E.,  {Gott} J.~Richard I.,  1972, \mn@doi [\apj] {10.1086/151605},
  \href {https://ui.adsabs.harvard.edu/abs/1972ApJ...176....1G} {176, 1}

\bibitem[\protect\citeauthoryear{{Hansen}, {Egli}, {Hollenstein}  \&
  {Salzmann}}{{Hansen} et~al.}{2005}]{2005NewA...10..379H}
{Hansen} S.~H.,  {Egli} D.,  {Hollenstein} L.,   {Salzmann} C.,  2005, \mn@doi
  [\na] {10.1016/j.newast.2005.01.005}, \href
  {https://ui.adsabs.harvard.edu/abs/2005NewA...10..379H} {10, 379}

\bibitem[\protect\citeauthoryear{{Johnston}, {Sigurdsson}  \&
  {Hernquist}}{{Johnston} et~al.}{1999}]{1999MNRAS.302..771J}
{Johnston} K.~V.,  {Sigurdsson} S.,   {Hernquist} L.,  1999, \mn@doi [\mnras]
  {10.1046/j.1365-8711.1999.02200.x}, \href
  {https://ui.adsabs.harvard.edu/abs/1999MNRAS.302..771J} {302, 771}

\bibitem[\protect\citeauthoryear{{Kelkar}, {Gray}, {Arag{\'o}n-Salamanca},
  {Rudnick}, {Jaff{\'e}}, {Jablonka}, {Moustakas}  \&
  {Milvang-Jensen}}{{Kelkar} et~al.}{2019}]{2019MNRAS.486..868K}
{Kelkar} K.,  {Gray} M.~E.,  {Arag{\'o}n-Salamanca} A.,  {Rudnick} G.,
  {Jaff{\'e}} Y.~L.,  {Jablonka} P.,  {Moustakas} J.,   {Milvang-Jensen} B.,
  2019, \mn@doi [\mnras] {10.1093/mnras/stz905}, \href
  {https://ui.adsabs.harvard.edu/abs/2019MNRAS.486..868K} {486, 868}

\bibitem[\protect\citeauthoryear{{Larson}}{{Larson}}{1974}]{1974MNRAS.169..229L}
{Larson} R.~B.,  1974, \mn@doi [\mnras] {10.1093/mnras/169.2.229}, \href
  {https://ui.adsabs.harvard.edu/abs/1974MNRAS.169..229L} {169, 229}

\bibitem[\protect\citeauthoryear{{Larson}, {Tinsley}  \& {Caldwell}}{{Larson}
  et~al.}{1980}]{1980ApJ...237..692L}
{Larson} R.~B.,  {Tinsley} B.~M.,   {Caldwell} C.~N.,  1980, \mn@doi [\apj]
  {10.1086/157917}, \href
  {https://ui.adsabs.harvard.edu/abs/1980ApJ...237..692L} {237, 692}

\bibitem[\protect\citeauthoryear{{Le Cam} \& {Yang}}{{Le Cam} \&
  {Yang}}{2012}]{le2012asymptotics}
{Le Cam} L.,  {Yang} G.~L.,  2012, Asymptotics in statistics: some basic
  concepts.
Springer Science \& Business Media

\bibitem[\protect\citeauthoryear{{Lopes}, {de Carvalho}, {Kohl-Moreira}  \&
  {Jones}}{{Lopes} et~al.}{2009a}]{2009MNRAS.392..135L}
{Lopes} P.~A.~A.,  {de Carvalho} R.~R.,  {Kohl-Moreira} J.~L.,   {Jones} C.,
  2009a, \mn@doi [\mnras] {10.1111/j.1365-2966.2008.13962.x}, \href
  {https://ui.adsabs.harvard.edu/abs/2009MNRAS.392..135L} {392, 135}

\bibitem[\protect\citeauthoryear{{Lopes}, {de Carvalho}, {Kohl-Moreira}  \&
  {Jones}}{{Lopes} et~al.}{2009b}]{2009MNRAS.399.2201L}
{Lopes} P.~A.~A.,  {de Carvalho} R.~R.,  {Kohl-Moreira} J.~L.,   {Jones} C.,
  2009b, \mn@doi [\mnras] {10.1111/j.1365-2966.2009.15425.x}, \href
  {https://ui.adsabs.harvard.edu/abs/2009MNRAS.399.2201L} {399, 2201}

\bibitem[\protect\citeauthoryear{{Lynden-Bell}}{{Lynden-Bell}}{1967}]{1967MNRAS.136..101L}
{Lynden-Bell} D.,  1967, \mn@doi [\mnras] {10.1093/mnras/136.1.101}, \href
  {https://ui.adsabs.harvard.edu/abs/1967MNRAS.136..101L} {136, 101}

\bibitem[\protect\citeauthoryear{{Mahajan}}{{Mahajan}}{2013}]{2013MNRAS.431L.117M}
{Mahajan} S.,  2013, \mn@doi [\mnras] {10.1093/mnrasl/slt021}, \href
  {https://ui.adsabs.harvard.edu/abs/2013MNRAS.431L.117M} {431, L117}

\bibitem[\protect\citeauthoryear{{Mahajan}, {Mamon}  \&
  {Raychaudhury}}{{Mahajan} et~al.}{2011}]{2011MNRAS.416.2882M}
{Mahajan} S.,  {Mamon} G.~A.,   {Raychaudhury} S.,  2011, \mn@doi [\mnras]
  {10.1111/j.1365-2966.2011.19236.x}, \href
  {https://ui.adsabs.harvard.edu/abs/2011MNRAS.416.2882M} {416, 2882}

\bibitem[\protect\citeauthoryear{{Martig}, {Bournaud}, {Teyssier}  \&
  {Dekel}}{{Martig} et~al.}{2009}]{2009ApJ...707..250M}
{Martig} M.,  {Bournaud} F.,  {Teyssier} R.,   {Dekel} A.,  2009, \mn@doi
  [\apj] {10.1088/0004-637X/707/1/250}, \href
  {https://ui.adsabs.harvard.edu/abs/2009ApJ...707..250M} {707, 250}

\bibitem[\protect\citeauthoryear{McKinney et~al.}{McKinney
  et~al.}{2010}]{mckinney2010data}
McKinney W.,  et~al., 2010, in Proceedings of the 9th Python in Science
  Conference. pp 51--56

\bibitem[\protect\citeauthoryear{Merrall \& Henriksen}{Merrall \&
  Henriksen}{2003}]{merrall2003relaxation}
Merrall T.~E.,  Henriksen R.~N.,  2003, The Astrophysical Journal, 595, 43

\bibitem[\protect\citeauthoryear{{Neistein}, {van den Bosch}  \&
  {Dekel}}{{Neistein} et~al.}{2006}]{2006MNRAS.372..933N}
{Neistein} E.,  {van den Bosch} F.~C.,   {Dekel} A.,  2006, \mn@doi [\mnras]
  {10.1111/j.1365-2966.2006.10918.x}, \href
  {https://ui.adsabs.harvard.edu/abs/2006MNRAS.372..933N} {372, 933}

\bibitem[\protect\citeauthoryear{{Ogorodnikov}}{{Ogorodnikov}}{1957}]{1957AZh....34..770O}
{Ogorodnikov} K.~F.,  1957, \azh, \href
  {https://ui.adsabs.harvard.edu/abs/1957AZh....34..770O} {34, 770}

\bibitem[\protect\citeauthoryear{{Oman}, {Hudson}  \& {Behroozi}}{{Oman}
  et~al.}{2013}]{2013MNRAS.431.2307O}
{Oman} K.~A.,  {Hudson} M.~J.,   {Behroozi} P.~S.,  2013, \mn@doi [\mnras]
  {10.1093/mnras/stt328}, \href
  {https://ui.adsabs.harvard.edu/abs/2013MNRAS.431.2307O} {431, 2307}

\bibitem[\protect\citeauthoryear{{Oman}, {Bah{\'e}}, {Healy}, {Hess}, {Hudson}
  \& {Verheijen}}{{Oman} et~al.}{2020}]{2020MNRAS.tmp.3623O}
{Oman} K.~A.,  {Bah{\'e}} Y.~M.,  {Healy} J.,  {Hess} K.~M.,  {Hudson} M.~J.,
  {Verheijen} M. A.~W.,  2020, \mn@doi [\mnras] {10.1093/mnras/staa3845}, \href
  {https://ui.adsabs.harvard.edu/abs/2020MNRAS.tmp.3623O} {}

\bibitem[\protect\citeauthoryear{{Paccagnella} et~al.,}{{Paccagnella}
  et~al.}{2016}]{2016ApJ...816L..25P}
{Paccagnella} A.,  et~al., 2016, \mn@doi [\apjl] {10.3847/2041-8205/816/2/L25},
  \href {https://ui.adsabs.harvard.edu/abs/2016ApJ...816L..25P} {816, L25}

\bibitem[\protect\citeauthoryear{{Park} \& {Choi}}{{Park} \&
  {Choi}}{2005}]{2005ApJ...635L..29P}
{Park} C.,  {Choi} Y.-Y.,  2005, \mn@doi [\apjl] {10.1086/499243}, \href
  {https://ui.adsabs.harvard.edu/abs/2005ApJ...635L..29P} {635, L29}

\bibitem[\protect\citeauthoryear{{Pasquali}, {Smith}, {Gallazzi}, {De Lucia},
  {Zibetti}, {Hirschmann}  \& {Yi}}{{Pasquali}
  et~al.}{2019}]{2019MNRAS.484.1702P}
{Pasquali} A.,  {Smith} R.,  {Gallazzi} A.,  {De Lucia} G.,  {Zibetti} S.,
  {Hirschmann} M.,   {Yi} S.~K.,  2019, \mn@doi [\mnras]
  {10.1093/mnras/sty3530}, \href
  {https://ui.adsabs.harvard.edu/abs/2019MNRAS.484.1702P} {484, 1702}

\bibitem[\protect\citeauthoryear{{Peng} et~al.,}{{Peng}
  et~al.}{2010}]{2010ApJ...721..193P}
{Peng} Y.-j.,  et~al., 2010, \mn@doi [\apj] {10.1088/0004-637X/721/1/193},
  \href {https://ui.adsabs.harvard.edu/abs/2010ApJ...721..193P} {721, 193}

\bibitem[\protect\citeauthoryear{{Read}, {Wilkinson}, {Evans}, {Gilmore}  \&
  {Kleyna}}{{Read} et~al.}{2006}]{2006MNRAS.366..429R}
{Read} J.~I.,  {Wilkinson} M.~I.,  {Evans} N.~W.,  {Gilmore} G.,   {Kleyna}
  J.~T.,  2006, \mn@doi [\mnras] {10.1111/j.1365-2966.2005.09861.x}, \href
  {https://ui.adsabs.harvard.edu/abs/2006MNRAS.366..429R} {366, 429}

\bibitem[\protect\citeauthoryear{Reynolds}{Reynolds}{2009}]{reynolds2009gaussian}
Reynolds D.~A.,  2009, Encyclopedia of biometrics, 741

\bibitem[\protect\citeauthoryear{{Rhee}, {Smith}, {Choi}, {Yi}, {Jaff{\'e}},
  {Candlish}  \& {S{\'a}nchez-J{\'a}nssen}}{{Rhee}
  et~al.}{2017}]{2017ApJ...843..128R}
{Rhee} J.,  {Smith} R.,  {Choi} H.,  {Yi} S.~K.,  {Jaff{\'e}} Y.,  {Candlish}
  G.,   {S{\'a}nchez-J{\'a}nssen} R.,  2017, \mn@doi [\apj]
  {10.3847/1538-4357/aa6d6c}, \href
  {https://ui.adsabs.harvard.edu/abs/2017ApJ...843..128R} {843, 128}

\bibitem[\protect\citeauthoryear{{Rhee}, {Smith}, {Choi}, {Contini}, {Jung},
  {Han}  \& {Yi}}{{Rhee} et~al.}{2020}]{2020ApJS..247...45R}
{Rhee} J.,  {Smith} R.,  {Choi} H.,  {Contini} E.,  {Jung} S.~L.,  {Han} S.,
  {Yi} S.~K.,  2020, \mn@doi [\apjs] {10.3847/1538-4365/ab7377}, \href
  {https://ui.adsabs.harvard.edu/abs/2020ApJS..247...45R} {247, 45}

\bibitem[\protect\citeauthoryear{{Ribeiro}, {Lopes}  \& {Trevisan}}{{Ribeiro}
  et~al.}{2010}]{2010MNRAS.409L.124R}
{Ribeiro} A.~L.~B.,  {Lopes} P.~A.~A.,   {Trevisan} M.,  2010, \mn@doi [\mnras]
  {10.1111/j.1745-3933.2010.00962.x}, \href
  {https://ui.adsabs.harvard.edu/abs/2010MNRAS.409L.124R} {409, L124}

\bibitem[\protect\citeauthoryear{{Ribeiro}, {de Carvalho}, {Trevisan},
  {Capelato}, {La Barbera}, {Lopes}  \& {Schilling}}{{Ribeiro}
  et~al.}{2013}]{2013MNRAS.434..784R}
{Ribeiro} A.~L.~B.,  {de Carvalho} R.~R.,  {Trevisan} M.,  {Capelato} H.~V.,
  {La Barbera} F.,  {Lopes} P.~A.~A.,   {Schilling} A.~C.,  2013, \mn@doi
  [\mnras] {10.1093/mnras/stt1071}, \href
  {https://ui.adsabs.harvard.edu/abs/2013MNRAS.434..784R} {434, 784}

\bibitem[\protect\citeauthoryear{{Roberts} \& {Haynes}}{{Roberts} \&
  {Haynes}}{1994}]{1994ARA&A..32..115R}
{Roberts} M.~S.,  {Haynes} M.~P.,  1994, \mn@doi [\araa]
  {10.1146/annurev.aa.32.090194.000555}, \href
  {https://ui.adsabs.harvard.edu/abs/1994ARA&A..32..115R} {32, 115}

\bibitem[\protect\citeauthoryear{{Roberts} \& {Parker}}{{Roberts} \&
  {Parker}}{2017}]{2017MNRAS.467.3268R}
{Roberts} I.~D.,  {Parker} L.~C.,  2017, \mn@doi [\mnras]
  {10.1093/mnras/stx317}, \href
  {https://ui.adsabs.harvard.edu/abs/2017MNRAS.467.3268R} {467, 3268}

\bibitem[\protect\citeauthoryear{{Roberts} \& {Parker}}{{Roberts} \&
  {Parker}}{2019}]{2019MNRAS.490..773R}
{Roberts} I.~D.,  {Parker} L.~C.,  2019, \mn@doi [\mnras]
  {10.1093/mnras/stz2666}, \href
  {https://ui.adsabs.harvard.edu/abs/2019MNRAS.490..773R} {490, 773}

\bibitem[\protect\citeauthoryear{{Roberts}, {Parker}, {Brown}, {Joshi},
  {Hlavacek-Larrondo}  \& {Wadsley}}{{Roberts}
  et~al.}{2019}]{2019ApJ...873...42R}
{Roberts} I.~D.,  {Parker} L.~C.,  {Brown} T.,  {Joshi} G.~D.,
  {Hlavacek-Larrondo} J.,   {Wadsley} J.,  2019, \mn@doi [\apj]
  {10.3847/1538-4357/ab04f7}, \href
  {https://ui.adsabs.harvard.edu/abs/2019ApJ...873...42R} {873, 42}

\bibitem[\protect\citeauthoryear{{Roediger} \& {Br{\"u}ggen}}{{Roediger} \&
  {Br{\"u}ggen}}{2007}]{2007MNRAS.380.1399R}
{Roediger} E.,  {Br{\"u}ggen} M.,  2007, \mn@doi [\mnras]
  {10.1111/j.1365-2966.2007.12241.x}, \href
  {https://ui.adsabs.harvard.edu/abs/2007MNRAS.380.1399R} {380, 1399}

\bibitem[\protect\citeauthoryear{{S{\'a}nchez-Bl{\'a}zquez}
  et~al.,}{{S{\'a}nchez-Bl{\'a}zquez} et~al.}{2006}]{2006MNRAS.371..703S}
{S{\'a}nchez-Bl{\'a}zquez} P.,  et~al., 2006, \mn@doi [\mnras]
  {10.1111/j.1365-2966.2006.10699.x}, \href
  {https://ui.adsabs.harvard.edu/abs/2006MNRAS.371..703S} {371, 703}

\bibitem[\protect\citeauthoryear{{Sarron}, {Adami}, {Durret}  \&
  {Laigle}}{{Sarron} et~al.}{2019}]{2019A&A...632A..49S}
{Sarron} F.,  {Adami} C.,  {Durret} F.,   {Laigle} C.,  2019, \mn@doi [\aap]
  {10.1051/0004-6361/201935394}, \href
  {https://ui.adsabs.harvard.edu/abs/2019A&A...632A..49S} {632, A49}

\bibitem[\protect\citeauthoryear{{Sazonova} et~al.,}{{Sazonova}
  et~al.}{2020}]{2020ApJ...899...85S}
{Sazonova} E.,  et~al., 2020, \mn@doi [\apj] {10.3847/1538-4357/aba42f}, \href
  {https://ui.adsabs.harvard.edu/abs/2020ApJ...899...85S} {899, 85}

\bibitem[\protect\citeauthoryear{{Schawinski} et~al.,}{{Schawinski}
  et~al.}{2014}]{2014MNRAS.440..889S}
{Schawinski} K.,  et~al., 2014, \mn@doi [\mnras] {10.1093/mnras/stu327}, \href
  {https://ui.adsabs.harvard.edu/abs/2014MNRAS.440..889S} {440, 889}

\bibitem[\protect\citeauthoryear{Schilling, Watkins  \& Watkins}{Schilling
  et~al.}{2002}]{SCH02}
Schilling M.~F.,  Watkins A.~E.,   Watkins W.,  2002, \mn@doi [The American
  Statistician] {10.1198/00031300265}, 56, 223

\bibitem[\protect\citeauthoryear{{Schuecker}, {B{\"o}hringer}, {Reiprich}  \&
  {Feretti}}{{Schuecker} et~al.}{2001}]{2001A&A...378..408S}
{Schuecker} P.,  {B{\"o}hringer} H.,  {Reiprich} T.~H.,   {Feretti} L.,  2001,
  \mn@doi [\aap] {10.1051/0004-6361:20011215}, \href
  {https://ui.adsabs.harvard.edu/abs/2001A&A...378..408S} {378, 408}

\bibitem[\protect\citeauthoryear{{Silverman}}{{Silverman}}{1986}]{1986desd.book.....S}
{Silverman} B.~W.,  1986, {Density estimation for statistics and data analysis}

\bibitem[\protect\citeauthoryear{{Springel} \& {Hernquist}}{{Springel} \&
  {Hernquist}}{2005}]{2005ApJ...622L...9S}
{Springel} V.,  {Hernquist} L.,  2005, \mn@doi [\apjl] {10.1086/429486}, \href
  {https://ui.adsabs.harvard.edu/abs/2005ApJ...622L...9S} {622, L9}

\bibitem[\protect\citeauthoryear{{Teyssier}, {Chapon}  \&
  {Bournaud}}{{Teyssier} et~al.}{2010}]{2010ApJ...720L.149T}
{Teyssier} R.,  {Chapon} D.,   {Bournaud} F.,  2010, \mn@doi [\apjl]
  {10.1088/2041-8205/720/2/L149}, \href
  {https://ui.adsabs.harvard.edu/abs/2010ApJ...720L.149T} {720, L149}

\bibitem[\protect\citeauthoryear{{Trevisan}, {Ferreras}, {de La Rosa}, {La
  Barbera}  \& {de Carvalho}}{{Trevisan} et~al.}{2012}]{2012ApJ...752L..27T}
{Trevisan} M.,  {Ferreras} I.,  {de La Rosa} I.~G.,  {La Barbera} F.,   {de
  Carvalho} R.~R.,  2012, \mn@doi [\apjl] {10.1088/2041-8205/752/2/L27}, \href
  {https://ui.adsabs.harvard.edu/abs/2012ApJ...752L..27T} {752, L27}

\bibitem[\protect\citeauthoryear{{Trussler}, {Maiolino}, {Maraston}, {Peng},
  {Thomas}, {Goddard}  \& {Lian}}{{Trussler}
  et~al.}{2020}]{2020MNRAS.491.5406T}
{Trussler} J.,  {Maiolino} R.,  {Maraston} C.,  {Peng} Y.,  {Thomas} D.,
  {Goddard} D.,   {Lian} J.,  2020, \mn@doi [\mnras] {10.1093/mnras/stz3286},
  \href {https://ui.adsabs.harvard.edu/abs/2020MNRAS.491.5406T} {491, 5406}

\bibitem[\protect\citeauthoryear{{Virtanen} et~al.,}{{Virtanen}
  et~al.}{2020}]{2020zndo....595738V}
{Virtanen} P.,  et~al., 2020, {scipy/scipy: SciPy 1.5.3},
  \mn@doi{10.5281/zenodo.595738}

\bibitem[\protect\citeauthoryear{Wetzel, Tinker  \& Conroy}{Wetzel
  et~al.}{2012a}]{wetzel2012galaxy}
Wetzel A.~R.,  Tinker J.~L.,   Conroy C.,  2012a, Monthly Notices of the Royal
  Astronomical Society, 424, 232

\bibitem[\protect\citeauthoryear{{Wetzel}, {Tinker}  \& {Conroy}}{{Wetzel}
  et~al.}{2012b}]{2012MNRAS.424..232W}
{Wetzel} A.~R.,  {Tinker} J.~L.,   {Conroy} C.,  2012b, \mn@doi [\mnras]
  {10.1111/j.1365-2966.2012.21188.x}, \href
  {https://ui.adsabs.harvard.edu/abs/2012MNRAS.424..232W} {424, 232}

\bibitem[\protect\citeauthoryear{{Wetzel}, {Tinker}, {Conroy}  \& {van den
  Bosch}}{{Wetzel} et~al.}{2013}]{2013MNRAS.432..336W}
{Wetzel} A.~R.,  {Tinker} J.~L.,  {Conroy} C.,   {van den Bosch} F.~C.,  2013,
  \mn@doi [\mnras] {10.1093/mnras/stt469}, \href
  {https://ui.adsabs.harvard.edu/abs/2013MNRAS.432..336W} {432, 336}

\bibitem[\protect\citeauthoryear{{Wojtak} \& {{\L}okas}}{{Wojtak} \&
  {{\L}okas}}{2010}]{2010MNRAS.408.2442W}
{Wojtak} R.,  {{\L}okas} E.~L.,  2010, \mn@doi [\mnras]
  {10.1111/j.1365-2966.2010.17297.x}, \href
  {https://ui.adsabs.harvard.edu/abs/2010MNRAS.408.2442W} {408, 2442}

\bibitem[\protect\citeauthoryear{{Yang}, {Mo}, {van den Bosch}, {Pasquali},
  {Li}  \& {Barden}}{{Yang} et~al.}{2007}]{2007ApJ...671..153Y}
{Yang} X.,  {Mo} H.~J.,  {van den Bosch} F.~C.,  {Pasquali} A.,  {Li} C.,
  {Barden} M.,  2007, \mn@doi [\apj] {10.1086/522027}, \href
  {https://ui.adsabs.harvard.edu/abs/2007ApJ...671..153Y} {671, 153}

\bibitem[\protect\citeauthoryear{{Zhang}, {Reiprich}, {Finoguenov}, {Hudson}
  \& {Sarazin}}{{Zhang} et~al.}{2009}]{2009ApJ...699.1178Z}
{Zhang} Y.-Y.,  {Reiprich} T.~H.,  {Finoguenov} A.,  {Hudson} D.~S.,
  {Sarazin} C.~L.,  2009, \mn@doi [\apj] {10.1088/0004-637X/699/2/1178}, \href
  {https://ui.adsabs.harvard.edu/abs/2009ApJ...699.1178Z} {699, 1178}

\bibitem[\protect\citeauthoryear{{Zu} \& {Mandelbaum}}{{Zu} \&
  {Mandelbaum}}{2016}]{2016MNRAS.457.4360Z}
{Zu} Y.,  {Mandelbaum} R.,  2016, \mn@doi [\mnras] {10.1093/mnras/stw221},
  \href {https://ui.adsabs.harvard.edu/abs/2016MNRAS.457.4360Z} {457, 4360}

\bibitem[\protect\citeauthoryear{{de Carvalho}, {Ribeiro}, {Stalder}, {Rosa},
  {Costa}  \& {Moura}}{{de Carvalho} et~al.}{2017}]{2017AJ....154...96D}
{de Carvalho} R.~R.,  {Ribeiro} A.~L.~B.,  {Stalder} D.~H.,  {Rosa} R.~R.,
  {Costa} A.~P.,   {Moura} T.~C.,  2017, \mn@doi [\aj]
  {10.3847/1538-3881/aa7f2b}, \href
  {https://ui.adsabs.harvard.edu/abs/2017AJ....154...96D} {154, 96}

\bibitem[\protect\citeauthoryear{{de Carvalho}, {Costa}, {Moura}  \&
  {Ribeiro}}{{de Carvalho} et~al.}{2019}]{2019MNRAS.487L..86D}
{de Carvalho} R.~R.,  {Costa} A.~P.,  {Moura} T.~C.,   {Ribeiro} A.~L.~B.,
  2019, \mn@doi [\mnras] {10.1093/mnrasl/slz084}, \href
  {https://ui.adsabs.harvard.edu/abs/2019MNRAS.487L..86D} {487, L86}

\bibitem[\protect\citeauthoryear{{de Vaucouleurs}}{{de
  Vaucouleurs}}{1963}]{1963ApJS....8...31D}
{de Vaucouleurs} G.,  1963, \mn@doi [\apjs] {10.1086/190084}, \href
  {https://ui.adsabs.harvard.edu/abs/1963ApJS....8...31D} {8, 31}

\bibitem[\protect\citeauthoryear{{van de Voort}, {Bah{\'e}}, {Bower}, {Correa},
  {Crain}, {Schaye}  \& {Theuns}}{{van de Voort}
  et~al.}{2017}]{2017MNRAS.466.3460V}
{van de Voort} F.,  {Bah{\'e}} Y.~M.,  {Bower} R.~G.,  {Correa} C.~A.,  {Crain}
  R.~A.,  {Schaye} J.,   {Theuns} T.,  2017, \mn@doi [\mnras]
  {10.1093/mnras/stw3356}, \href
  {https://ui.adsabs.harvard.edu/abs/2017MNRAS.466.3460V} {466, 3460}

\bibitem[\protect\citeauthoryear{{van der Walt}, {Colbert}  \&
  {Varoquaux}}{{van der Walt} et~al.}{2011}]{2011CSE....13b..22V}
{van der Walt} S.,  {Colbert} S.~C.,   {Varoquaux} G.,  2011, \mn@doi
  [Computing in Science and Engineering] {10.1109/MCSE.2011.37}, \href
  {https://ui.adsabs.harvard.edu/abs/2011CSE....13b..22V} {13, 22}

\makeatother
\end{thebibliography}




\appendix

\section{Comparison Between Rhee Regions vs. Pasquali's New Zones}
\begin{figure*}
    \centering
    \includegraphics[width = \textwidth]{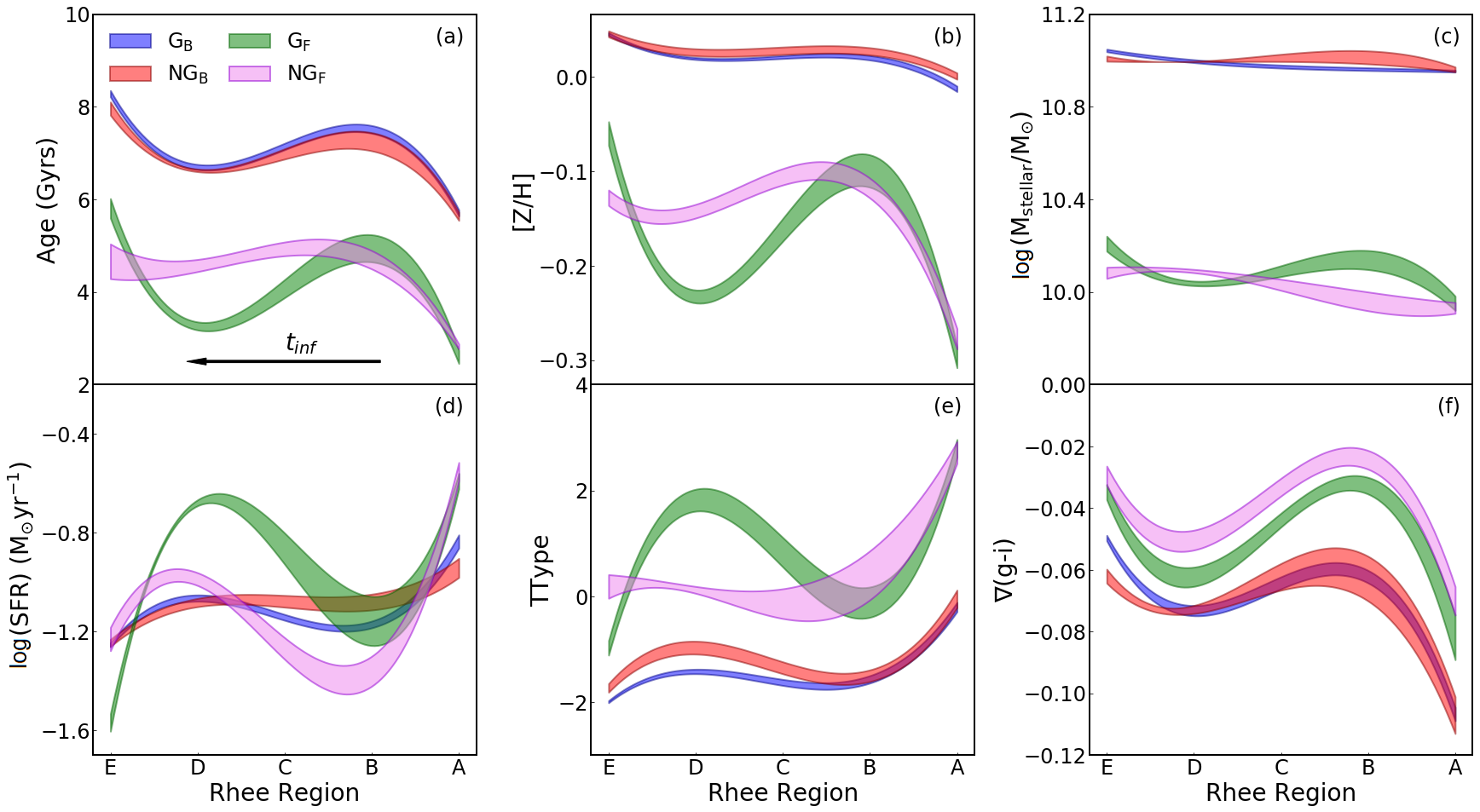}
    \caption{Same as Fig.~\ref{fig:tinfall_relation}, but using Rhee Regions (instead of PNZ) in order to map the location on the PPS and $\rm t_{inf}$. Colors follow the same schema of Fig. \ref{fig:tinfall_relation}. The curves and their width represent a spline fitting and the associated variance, respectively.}
    \label{fig:rhee_relation}
\end{figure*}

There are different ways to separate galaxies in projected phase
space. In this work we slice the PPS in PNZs \citep{2019MNRAS.484.1702P}, which
divide galaxies according to their infall time. This choice, however,
is not unique. The definition of PNZs is restricted to cluster-centric
distances within $R_{200}$. To explore the outer areas of clusters, we also 
adopt the ``Rhee Regions'' (R17) as a second way
of dividing the PPS tracing galaxies infall time. This second way of
discretizing the PPS is based on a probabilistic approach. Therefore, there is no direct relation between Rhee Regions and PNZs. In Fig.~\ref{fig:rhee_relation}
we show the relation between Rhee Region and galaxy properties. 
Examining Figs~\ref{fig:tinfall_relation} and \ref{fig:rhee_relation} we find that our analysis with Rhee Regions or PNZs lead to quite similar global trends. Similarly to the PNZs case, we find that the most striking differences are in the faint regime. Additionally, the similarity extends to the trends we observe. From panels (a) and (b), we note that F galaxies in regions D and C of NG clusters are older and more metal-rich than its counterpart in G systems. In panel (c), we find that there are no significant differences between stellar mass of F galaxies in G or NG systems. Regarding SFR (panel d), comparison shows that faint galaxies in regions D, C and B of G clusters are more star forming than those in NG systems. This is also related to morphology (panel e), for which F galaxies in regions D and C of NG clusters have higher TType values (closer to elliptical). Finally, in panel (f) we note that F galaxies in NG clusters have shallower (closer to 0) color gradients than in G systems. However, We highlight that there are less Rhee Regions than PNZs, which leads to a more "curvy" shaped curves due to the adopted procedure. This paper performs most of
the analysis using the PNZs and, when the $\rm R_{200}$ limitation plays a
major role (Section~\ref{sec:infall_rate_estimate}, for instance), we
switch to the alternative scheme.


\bsp	
\label{lastpage}
\end{document}